\documentclass[twocolumn,showpacs,preprintnumbers,amsmath,amssymb,prb]{revtex4}

\usepackage{graphicx}
\usepackage{dcolumn}
\usepackage{bm}

\draft

\begin{document}
\title{Out-of-equilibrium dynamical fluctuations in glassy systems}
\author{C. Chamon}
    \email{chamon@buphy.bu.edu}
    \affiliation{Department of Physics, Boston University, USA\\
     590 Commonwealth Avenue, Boston MA 02215 USA}
\author{P. Charbonneau}
    \affiliation{Department of Chemistry and Chemical Biology,
    Harvard University\\
    12 Oxford Street, Cambridge MA 02138 USA
    }
\author{L. F. Cugliandolo}
    \email{leticia@lpt.ens.fr}
    \affiliation{Laboratoire de Physique Th\'eorique, Ecole Normale
    Sup\'erieure,\\
    24 rue Lhomond, 75231 Paris Cedex 05, France \\
    Laboratoire de Physique Th\'eorique et Hautes Energies, Jussieu, \\
    4 Place Jussieu 75252 Paris Cedex 05, France
    }
    \altaffiliation[Research Associate at ]{The Abdus Salam ICTP,
      Trieste, Italy}
\author{D. R. Reichman}
    \email{reichman@chem.harvard.edu}
    \affiliation{Department of Chemistry and Chemical Biology,
    Harvard University\\
    12 Oxford Street, Cambridge, MA  02138 USA
    }
\author{M. Sellitto}
    \email{sellitto@ictp.trieste.it}
    \affiliation{The Abdus Salam International Centre for Theoretical
      Physics \\
    Strada Costiera 11, 34100 Trieste, Italy}

\date{\today}
\begin{abstract}
  In this paper we extend the earlier treatment of out-of-equilibrium
  mesoscopic fluctuations in glassy systems in several significant
  ways.  First, via extensive simulations, we demonstrate that models
  of glassy behavior without quenched disorder display scalings of the
  probability of local two-time correlators that are qualitatively
  similar to that of models with short-ranged quenched interactions.
  The key ingredient for such scaling properties is shown to be the
  development of a critical-like dynamical correlation length, and not
  other microscopic details.  This robust data collapse may be
  described in terms of a time-evolving Gumbel-like distribution.  We
  develop a theory to describe both the form and evolution of these
  distributions based on a effective sigma-model approach.
\end{abstract}
\maketitle

\section{Introduction}

The dynamics of glassy systems is extremely slow, both near and far
from equilibrium~\cite{review-glasses}.  Until recently, experiments,
numerical studies and theory focused on the study of averaged global,
or bulk, dynamical quantities that demonstrate the nonequilibrium
character of the glassy state. Interest is now shifting to the study
of local quantities embodied in intermittent fluctuations and dynamic
heterogeneities.  The recent development of powerful experimental
techniques~\cite{Ediger,Kegel,Weitz,Weeks1,Weeks2,Nathan,Miller,Sergio2,Luca2}
and the use of extensive numerical simulations have started to yield
valuable information about the dynamics of supercooled
liquids~\cite{Laird,2d,Donati99,Heuer,Johnson,Yamamoto,Starr02,Ogli,Vollmayr}
and glassy systems~\cite{Castilloetal,Castilloetal2} at a mesoscopic
scale.  A detailed analysis of temporal fluctuations in the evolution
of global observables allows one to distinguish between intermittent
and continuous dynamics, the former being associated with sudden,
large rearrangements~\cite{Weeks2,Sergio2,Luca2}. In addition, the
study of run-to-run fluctuations in systems of mesoscopic size yields
complementary information~\cite{Castilloetal2}.  The insight obtained
from these studies is useful in identifying the most relevant
rearrangements that take place during evolution and thus may lead to a
better understanding of the mechanism leading to glassy arrest.

The behavior of global dynamic correlations in macroscopic glassy
systems is rather well-described analytically with an extension of the
mode-coupling theory to the low temperature
regime~\cite{review-Bouchaud,Cugliandolo}. This approach is also
attractive since, via its connection to mean-field models with
quenched disorder, it allows for an interpretation of dynamic arrest
and slow dynamics in terms of the geometric properties of a
free-energy landscape~\cite{Cuku93,Bocukume,free-energy,inherent}.
However, it is not clear how to describe temporal and spatial
fluctuations within this approach.  Another picture of glassy dynamics
is based on kinetically constrained models with trivial static
interactions~\cite{FA,KoAn,JaKr} and is successful in describing aging
effects and other glassy features~\cite{aging-kinetic} as has been
recently reviewed~\cite{reviews-kinetic}. Moreover, dynamic
heterogeneities also find a natural interpretation within these
models~\cite{Harrowell,Chandler}.  The connection between the
microscopic interacting real system and the non-interacting
kinetically constrained model is however phenomenological, as the
facilitation rules are based on a coarse-grained description of the
liquid dynamics.  Other phenomenological descriptions do not even have
a real-space interpretation. Thus, one of the more interesting and
open theoretical problems at present is how to derive the observed
properties of medium-size and large-scale fluctuations that are
possibly associated with cooperative and correlated rearrangements of
the microscopic constituents (particles, spins, etc.) from a
microscopic model~\cite{Chandler}.

In this paper we focus on particular aspects of the non-equilibrium
dynamics near or below the onset of glassy arrest. Some of our
arguments and numerical analysis can be simplified to capture the
(stationary) slow dynamics of supercooled liquids as a particular
case.  The purpose of this paper is three-fold. First, we test whether
the scenario proposed in
Refs.~\cite{Castilloetal,Castilloetal2,Chamonetal} for spatially
heterogeneous glassy dynamics applies to models without quenched
disorder such as kinetically constrained spin models.  Second, we show
that several features of the probability distributions functions
({\textsc{pdf}}'s) of fluctuating two-time correlations, namely time
scalings and functional forms, are common to many glassy systems.
Third, we discuss an analytic approach developed to describe the
scaling form and time evolution of these {\textsc{pdf}}'s.  This
approach is based on the arguments put forward in
\cite{Castilloetal,Chamonetal} and on an analogy with the study of the
order-parameter fluctuations in critical phenomena~\cite{book}, where
one can describe universality classes in terms of the {\textsc{pdf}}'s
of global fluctuations.  The latter is similar in spirit to the work
of R\'acz and collaborators on interfacial growth
problems~\cite{Racz}, and Bramwell and coworkers on the {\textsc{xy}}
model~\cite{Peter}.

Even if, strictly speaking, there is no liquid-glass dynamic phase
transition, glassy dynamics present some aspects of critical dynamics.
Among others, a dynamic correlation length defined from the
fluctuations of the local two-time correlations shows an apparent
divergence at long
times~\cite{Castilloetal2,Sharon,Silvio,Whitelam,Biroli}.  In the
critical situation, one expects that the symmetries of the problem
constrain the form and scaling of the {\textsc{pdf}}'s of fluctuating
dynamical quantities.  A natural way to attempt an analytic approach
to these fluctuations is then to construct an effective sigma model
that respects these symmetries.  This approach has been advocated in
\cite{Castilloetal,Castilloetal2,Chamonetal}. Here we extend it to
propose a model that describes not only the time-scaling of the
{\textsc{pdf}}'s but also their functional form.

The plan of the paper is as follows. In Section~\ref{sect2} we
summarize the main properties of glassy dynamics as displayed by the
decay of global correlation functions. In Section~\ref{exps} we
present a short review of experiments that focus on the measurement of
dynamic fluctuations. In Section~\ref{models} we introduce the models
that we study numerically; these are a modification of the
Fredrickson-Andersen ({\textsc{fa}}) model due to Graham {\it et
  al.}~\cite{Graham} and the Edwards-Anderson ({\textsc{ea}})
spin-glass~\cite{EA}.  In Section~\ref{fluctuations} we present the
{\textsc{pdf}}'s for fluctuating two-time correlations.  We
demonstrate that very similar {\textsc{pdf}}'s are obtained from
facilitated Ising models without quenched random interactions and the
disordered spin-glass problem. These forms also resemble strongly the
experimental data of Cipelletti {\it et al.}~\cite{Luca2} for
colloidal suspensions. We describe these {\textsc{pdf}}'s with an
evolving Gumbel-like form that we later argue can be justified from
the sigma-model approach.  In Section~\ref{analytic} we introduce the
theoretical modeling of these two-time dependent {\textsc{pdf}}'s.
Finally, in Section~\ref{conclusions} we present our conclusions.
Appendix~\ref{gumbel} discusses the definition of a Gumbel
distribution and some of its properties. The aim of
Appendix~\ref{app:toy} is to identify which features of the {\sc pdf}s
are trully due to glassy dynamics. To this end we analyse the
stochastic dynamics (with no facilitation rule) of a finite system of
independent spins on a lattice.

\section{Nonequilibrium relaxation in glassy systems}
\label{sect2}

Glassy systems show slow nonexponential dynamical relaxation.  This is
demonstrated by the extremely sluggish decay of one-time observables
like the energy density, the structure factor, etc. In many
experimental cases, one-time observables can be considered to be
approximately constant within the time window explored. Thus,
observables that depend on only one time are not sufficient to
characterize the non-equilibrium relaxation of a glassy system in its
full richness; they do not signal one of the hallmarks of glassy
systems, namely aging phenomena.  Aging is partially described by the
decay of two-time global correlation functions:
\begin{eqnarray}
  C(t,t_w) &\equiv& {\cal N}^{-1} \langle O(t) O(t_w) \rangle
\label{global-corr}
\end{eqnarray}
with $O(t)$ the global observable of interest and ${\cal N}$ a
two-time dependent normalization that enforces $C(t,t)=1$. The angular
brackets denote an average over different thermal histories of the
system. In a model with quenched disorder, e.g. the {\textsc{ea}}
spin-glass, one averages the global correlation over different
realizations of the interactions.  Sometimes a coarse-graining in time
using a short time window around the observation instant is also
implemented.  It is important to note that for a sufficiently large
system, the quantity
\begin{equation}
  O(t) O(t_w)/{\cal N}
\label{global}
\end{equation}
does not fluctuate and the result obtained for a single thermal
history (and a single realization of disorder) coincides with the
averaged one, as long as times are long, but not so long as to exhibit
recovery of equilibrium behavior.

As already mentioned, one-time observables evolve very slowly in the
glassy phase of real systems. Still, some models used to capture
glassy dynamics have a faster evolution of one-time quantities.  In
these cases, it is convenient to study the evolution of the {\it
  connected correlations} defined as in Eq.~(\ref{global-corr}) where
\begin{equation}
O(t) \to O(t) - \langle O(t) \rangle
\; .
\end{equation}

By the very definition of aging, older systems relax in a slower
manner than younger ones. One defines the age of a system as the time
spent in the phase under study. A way to characterize the aging
properties is to monitor the time evolution of the two-time
correlation defined in Eq.~(\ref{global-corr}) or the two-time
function in Eq.~(\ref{global}). Well below $T_g$ (conventionally
defined as the value at which the relaxation time reaches 100s) the
system is in a ``phase'' that we call ``glassy''. At such low
temperatures the correlations depend explicitly on $t_w$ and, within
the experimentally accessible time-window, do not show a tendency
towards decay to equilibrium behavior, that is to say an independence
from $t_w$.  Close to glassy arrest one observes ``interrupted
aging,'' namely a dependence on the age of the system until
observations reach the equilibration time ($t_w > t_{\textsc{eq}}$)
where the dynamics crosses over to an equilibrium one.

The correlations of Eq.~(\ref{global-corr}) have been studied
experimentally in a number of glassy materials.  In spin systems the
natural choice for $O(t)$ is the fluctuation of the global
magnetization~\cite{Didier}, while in colloidal suspensions it is the
intensity of scattered light~\cite{Luca2,Bonn,Virgile,Luca1}. Many
numerical studies of a large variety of models have focused on the
analysis of the closely related intermediate scattering function.  On
the analytical side, the study of mean-field disordered models such as
the $p$-spin model yielded several features of the aging behavior seen
in simulations and experiments~\cite{Cugliandolo,Cuku93}.  Connected
global correlations in several kinetically constrained spin systems
have also been studied yielding similar results~\cite{aging-kinetic,
  reviews-kinetic}.

In structural glasses, a pictorial explanation of aging may be put
forward by imagining that each particle sees a cage made of its
neighbors.  When $\tau$ is short compared to a characteristic time
$\tau_0(t_w)$, each particle rattles within its cage, decorrelation is
characterized only by thermal fluctuations, and correlations decay in
a stationary manner from unity at equal times to a plateau,
$q_{\textsc{ea}}(T)$, seen on a logarithmic scale.  When $\tau$
increases, the motion of the particles destroys the cages and the
structure of the system relaxes.  The waiting-time dependence implies
that the cages become stiffer as time evolves. The motion of a tagged
particle observed with confocal microscopy demonstrated this
scenario~\cite{Weeks2}.  Upon examination of the value of the
displacement reached at the maximum separation time, however, one
notices that motion occurs only on length scales of the order of the
radius of a particle.  This means that the structural relaxation
occurs without large scale structural rearrangement and with
essentially no exchange of particles within the bulk of the sample.
The same observation applies to the dynamics of granular matter under
external forcing~\cite{Pouliquen}.

The existence of a well-defined cage translates into the appearance of
a plateau in the decay of the two-time correlations.  The presence of
a plateau in the correlation function means that there is a
macroscopic time interval in which the correlation (when plotted in
log-log scale) can be approximated by a (non-zero) constant.  In this
case the two-time correlation can be scaled in a satisfactory way by
using a separation in fast and slow
regimes~\cite{Cugliandolo,Cuku93,Cuku2}:
\begin{eqnarray}
  C(t,t_w) &=& C_{\textsc{f}}(t-t_w) + C_{\textsc{s}}(t,t_w) \nonumber
  \\ &=& f_{\textsc{f}}
  \left(\frac{h_{\textsc{f}}(t)}{h_{\textsc{f}}(t_w)} \right) +
  C_{\textsc{s}}(t,t_w)
\label{global-C}
\end{eqnarray}
with $h_{\textsc{f}}(t) = {\rm e}^{t/\tau_{\textsc{f}}}$ and
$\tau_{\textsc{f}}$ a characteristic time scale. The first term
describes the stationary approach to the plateau, the second term the
$t_w$-dependent departure from it.  The light-scattering study of a
colloidal system of Viasnoff {\it et al.}~\cite{Virgile} shows a very
clear plateau separating the two regimes.  However, a clear-cut
plateau is not always visible in numerical simulations and
experiments.

In most cases studied so far the second (aging) decay is characterized
by a similar form
\begin{equation}
  C_{\textsc{s}}(t,t_w) = f_{\textsc{s}}
  \left(\frac{h_{\textsc{s}}(t)}{h_{\textsc{s}}(t_w)}\right) \; ,
\label{slow-C}
\end{equation}
with $h_{\textsc{s}}(t)$ a system-dependent monotonic function.  A
typical and well-understood example in which the slow part of the
decay is represented by this form is the case of systems undergoing
domain growth. In this case, for long and well-separated times,
$h_{\textsc{s}}(t)$ is the typical size of the domains at time $t$.
Other examples include the relaxation of the incoherent scattering
function in Lennard-Jones mixtures and the result of light-scattering
measurements in colloidal suspensions~\cite{Luca2,Virgile,Luca1}.
Monte Carlo data for the two-time spin-spin correlations in the $3d$
{\textsc{ea}} model can be described with such a two-step relaxation
and $h_{\textsc{s}}(t)=t$~\cite{Marco}.  Similarly, the magnetization
correlations in the insulating spin-glass studied by H\'erisson and
Ocio~\cite{Didier} are described by a two-step relaxation.  In this
study, an enhanced power law, $h_{\textsc{s}}(t)= {\rm e}^{\ln^a
  t/t_0}$, yields the best fit of the experimental data. We shall see
below that, for the times that one can reach with a numerical
simulation, the kinetically constrained models defined in
Sect.~\ref{models} either do not show a plateau in their connected
correlations, or show several plateaus. Moreover, it is very difficult
to identify a single function $h_{\textsc{s}}(t)$ that describes the
slow decay.  This might be due to the fact that the slow decay
involves several correlation scales that are still not well separated
during numerical times.

\section{Heterogeneous dynamics}
\label{exps}

The existence of dynamic heterogeneities in supercooled liquids and
glasses has been suggested on the basis of experiments performed using
many techniques~\cite{Ediger}; two of them are the confocal microscopy
and light scattering. We discuss these results in some detail since we
shall compare our numerical simulations to them.

Confocal microscopy allows one to reconstruct the trajectory of each
particle in a colloidal suspensions made of several thousand
particles~\cite{Kegel,Weitz}. Kegel and van Blaaderen~\cite{Kegel}
found a non-Gaussian distribution of particle displacements in a dense
system of colloidal hard spheres and they associated this large
distribution to the presence of dynamical heterogeneities. Later,
Courtland and Weeks found that mobile (and immobile) particles cluster
in regions of lower (higher) density than average in the glassy phase
of a colloidal system~\cite{Weeks2}.

The ``time-resolved light-scattering technique''~\cite{Luca2} has been
developed with the aim of testing the fluctuations in time of the
intensity-intensity correlations in samples of medium size.  The
existence of intermittency in measurements of {\it global} quantities
may be another manifestation of heterogeneous dynamics.  Temporal
fluctuations in colloidal suspensions have been studied in this way.
Interestingly enough, Cipelletti's group showed that aging samples,
such as concentrated colloidal gels, have an intermittent dynamics
leading to negatively skewed distributions of the two-time
intensity-intensity correlations when the time lag is shorter than the
relaxation time and Gaussian distributions when the time lag
approaches the relaxation time.  They also showed that simpler systems
with no glassy features, such as Brownian colloidal samples, have a
continuous relaxation with a Gaussian distribution of the same
quantities for all time lags explored. The same trend of time resolved
{\textsc{pdf}}'s of particle displacements was demonstrated by
Courtland and Weeks using confocal microscopy~\cite{Weeks2}.  It is
interesting to note that these experiments show that significant
fluctuations occur even if the time lag $\tau$ is very short compared
to $t_w$.

Evidence for intermittency in the {\it global} voltage signal noise in
a solid-like colloidal glass and a polymer glass has also been
presented recently~\cite{Sergio2}. While the samples are out of
equilibrium the voltage time series has bursts with very large
amplitudes.  These bursts become rarer with the sample age in the
sense that their amplitude decreases and the time elapsed between two
consecutive bursts increases. When aging stops the bursts disappear
from the noise signal.  The waiting-time dependent {\textsc{pdf}},
$\rho(V)$, constructed by dividing time in intervals of duration
$\tau$ is clearly negatively skewed at short times and progressively
approaches a Gaussian distribution when the age of the system
increases~\cite{Sergio2}.

On the numerical side, most studies have focused on the supercooled
liquid phase of binary
mixtures~\cite{Laird,2d,Donati99,Heuer,Johnson,Yamamoto} and polymer
melts~\cite{Starr02} and the heterogeneities have been studied by
tagging each particle and classifying them according to their mobility
during a selected time interval.  Some recent papers treat the glassy
phase of the Lennard-Jones system in a similar way~\cite{Vollmayr}.
Following the dynamics of individual particles is very useful to get
an intuitive understanding of the evolution of the system, but it is
hard to use as a direct input in a theoretical approach. In order to
develop such a description, it is much more convenient to use the
coarse-grained two-time quantities~\cite{Castilloetal,Chamonetal} that
we shall define in Sect.~\ref{fluctuations}.  In this setting, a
heterogeneity is a region in the sample that relaxes very differently
from the bulk.  Clearly, the heterogeneities are not static and they
appear and disappear as dynamic fluctuations.

\section{The models}
\label{models}

We shall focus on several spin models that describe and capture
different aspects of glassy systems. We define them below.

\subsection{Kinetically constrained model}

\subsubsection{A variation on the Fredrickson-Andersen model}

The first model we study is a kinetically constrained spin system with
trivial equilibrium properties.  It is a simple variation of the
Fredrickson-Andersen ({\textsc{fa}}) model~\cite{FA} due to Graham {\it et
al.}~\cite{Graham}.  $N$ Ising spins, $s_i=\pm 1$, $i=1,\dots,N$,
occupy the vertices of a regular cubic lattice in $d$ dimensions. The
energy is given by
\begin{equation}
  E = - \frac{h}{2} \sum_{i=1}^N s_i \; .
\label{kc-energy}
\end{equation}
Thus there are no static interactions between the spins but an
external field of strength $h$ is applied. The system is in contact
with a heat bath at temperature $T$ that, for convenience, we shall
measure in units of the external field $h$.  This model does not have
a static transition and at equilibrium the averaged magnetization,
$m=\sum_i \langle s_i \rangle/N $, is given by
\begin{equation}
  m = \tanh \frac{h}{2 \, k_{\textsc B}T}  \;.
\label{kc-magnetization}
\end{equation}
The non-trivial evolution of this non-interacting system arises due to
the constrained dynamics imposed on the spins. At each time step a
spin, $s_k$, that is chosen at random among the ensemble is allowed to
flip with a generalized heat bath rule. If at least $p$ of its $z=2d$
neighbors point down, {\it i.e.} opposite to the preferred direction,
the move is accepted with unit probability if $\Delta E<0$ and with
probability ${\rm e}^{-\Delta E/k_{\textsc B}T}$ if $\Delta E>0$.
$\Delta E=E_f-E_i$, with $E_f$ the total energy after the move and
$E_i$ the total energy before the move. In this case, $\Delta
E=(-s_k^f+s_k^i)/2= s_k^i$.  Otherwise the attempted flip is rejected.
The parameter $p$ defines the facilitated rule. For instance, a
three-site rule requires a minimum of $3$ neighbors to oppose the
field direction. As usual, time units are defined via the number of
attempts to flip randomly chosen spins in the sample.  The
facilitation rule satisfies detailed balance.  Thus the system evolves
towards the equilibrium configuration, although it may do so in a very
slow manner. The larger the value of $p$ the slower the resulting
dynamics.

Typically, one is interested in following the relaxational dynamics of
a random initial condition in which each spin points in the up or down
direction with probability one half. This mimics a rapid quench from
infinite temperature to the working temperature $T$.  The dynamics
tend to order the system in the up direction. However, as time passes,
it is harder to choose a spin with sufficient number of neighbors
pointing down. This induces a dynamic slowing down, and below $T\sim
0.5$ (in $d=3$ and using $p=3$) or $T\sim 0.7$ (in $d=2$ and using
$p=2$) the system does not manage to get close to the equilibrium
configuration. The out-of-equilibrium relaxation below this crossover
temperature resembles that of supercooled liquids~\cite{Mauro}.

Importantly, these models do not contain quenched disorder.  The
absence of quenched disorder and the local, non-mean-field nature of
the relaxation strongly suggests that such models are relevant for
understanding dynamics of supercooled liquids.  The existence or
absence of a genuine glass transition is not important for our
purposes. Indeed, as it has been recently shown rigorously is some
cases these models do not have a dynamic phase transition in the
thermodynamic limit~\cite{Cristina}.  Nevertheless, the evolution of a
large finite system at long but finite times shares some keynote
properties with many glassy systems, including systems that have a
finite temperature dynamic transition, such as spin-glass models that
undergo one-step replica symmetry breaking.  Thus, the study of
facilitated models in the context presented here is of even broader
importance than just the modeling of supercooled liquids.

\subsection{The Edwards-Anderson spin glass}

A disordered model that we shall briefly discuss here is the standard
non-mean-field model for spin-glasses, the finite dimensional
Edwards-Anderson model~\cite{EA}. The energy function is highly
non-trivial in this case,
\begin{equation}
  E_J = - \sum_{\langle ij\rangle } J_{ij} s_i s_j \;,
\label{ea-energy}
\end{equation}
where the spins are Ising variables that occupy the vertices of a
cubic $d$-dimensional lattice, the sum runs over nearest-neighbors,
and the coupling strengths are random and chosen from a bimodal or
Gaussian probability distribution with zero mean and variance
$[J_{ij}^2]=1/(2z)$. The square brackets denote an average over the
random exchanges. This model is said to be {\it disordered} due to the
quenched randomness introduced via the exchanges. The static
properties are still under debate. It is believed that in $3d$ the
Edwards-Anderson model undergoes a static phase transition from a
paramagnetic to a spin-glass phase at a finite value of $T$, which
depends on the type of disorder used, but there is no consensus about
the nature and properties of the static spin-glass
phase~\cite{review-spinglass}.  Dynamically, one may also follow the
evolution of spin configurations from a random initial condition. The
dynamics is, in this case, usual Monte Carlo with the heat bath rule
determined by the energy function (\ref{ea-energy}). The evolution of
a sufficiently large system at low temperatures occurs out of
equilibrium.

The local mesoscopic fluctuations of this model have been studied
analytically and numerically
in~\cite{Castilloetal,Castilloetal2,Chamonetal}.  We shall not
reproduce these results here.  In this manuscript we merely augment
the results presented in~\cite{Castilloetal,Castilloetal2,Chamonetal}
to test the properties of fluctuations that have not been discussed in
previous publications.

\section{Fluctuations}
\label{fluctuations}

In this section we introduce the quantities we shall focus on and we
discuss the behavior of their {\textsc{pdf}}'s as found for the
kinetically constrained models and the $3d$ {\textsc{ea}} spin-glass.

\subsection{Spatial fluctuations}

Local fluctuations in an unperturbed system are monitored by studying
\begin{eqnarray}
  C^o_r(t,t_w) &\equiv & \frac{1}{V_r} \sum_{x \in V_r} O(x,t)
  O(x,t_w) - \\\nonumber &&\frac{1}{V_r} \sum_{x \in V_r} O(x,t) \;
  \frac{1}{V_r} \sum_{x'\in V_r} O(x',t_w)
\label{local-corr0}
\end{eqnarray}
where $V_r=\ell^d$ is a coarse-graining volume centered at the
position $r$. This is a local connected function that evolves from a
time-dependent value at equal times to zero at diverging time
separations. Since it is convenient to work with a normalized two-time
quantity at equal times, we define
\begin{eqnarray}
  C_r(t,t_w) \equiv \frac{C_r^o(t,t_w)}{\sqrt{C_r^o(t,t)
  C_r^o(t_w,t_w)}} \;.
\label{local-corr1}
\end{eqnarray}
With this definition the normalization itself is site-dependent and
might introduce spurious fluctuations. Another possibility is to
normalize the local correlation with the global connected one
\begin{equation}
  C_r(t,t_w) \equiv \frac{C_r^o(t,t_w)}{\sqrt{C(t,t)
  C(t_w,t_w)}} \;.
\label{local-corr2}
\end{equation}
This normalization bounds the global correlation to be less than unity
but it does not bound the local one; in particular, it does not
enforce the local correlation to be one at equal times.  In what
follows, we present results obtained using definition
\ref{local-corr2} only.

For lattice spin models, the choice is
\begin{eqnarray}
  C^o_r(t,t_w) &\equiv& \frac{1}{V_r} \sum_{i\in V_r} s_i(t)
  s_i(t_w)-\\\nonumber &&\frac{1}{V_r} \sum_{i\in V_r} s_i(t)
  \frac{1}{V_r} \sum_{j\in V_r} s_j(t_w)
\label{local-corr}
\end{eqnarray}
and its normalized counterpart.  For the {\textsc{ea}}
model one easily verifies that, when the coarse-graining volume is
sufficiently large, the local magnetization vanishes, $V^{-1}_r
\sum_{i\in V_r} s_i(t_w)=0$, and there is no need to use the connected
functions above.  For the modified {\textsc{fa}} model, the local
magnetization does not vanish and we shall work with connected
functions.  In none of these expressions is there any averaging, apart
from the coarse-graining in space.

With the local observables one can define a two-time dependent
correlation length $\xi(t,t_w)$ via~\cite{Castilloetal2}
\begin{eqnarray*}
&& A(\Delta; t,t_w )
\equiv
\\
&&
\;\;
\frac{1}{V} \! \int d^d x \,\,
O(\vec x,t) O(\vec x,t_w) O(\vec x+\vec \Delta,t) O(\vec x+\vec \Delta,t_w)
\end{eqnarray*}
and the normalized quantity
\begin{eqnarray*}
B(\Delta, t,t_w) &\equiv&
\frac{A(\Delta; t,t_w)-A(\Delta\to\infty; t,t_w )}
{A(\Delta=0; t,t_w)-A(\Delta\to\infty; t,t_w )}
\\
&\sim&
{\rm e}^{-\Delta/\xi(t,t_w)}
\;.
\end{eqnarray*}
(We have made explicit the vectorial character of the positions $\vec
x$ in $d>1$. In the case of spin problems it is convenient to use
$O(\vec x,t) = \overline s_i(t)$ where the overline indicates a
coarse-graining in time over a time-window centered on $t$.)

Depending on the relative value between the linear size of the
coarse-graining volume, $\ell$, and the two-time dependent correlation
length, $\xi$, we expect either:

{\it i.} If $\ell$ is not much larger than $\xi(t,t_w)$ different
cells in the system behave rather differently and the histogram of the
quantities (\ref{local-corr0}) leads to a non-trivial {\textsc{pdf}}
whose form depends on the two times chosen, $t_w$ and $t$.

{\it ii.}  If $\ell \gg \xi(t,t_w)$ the coarse-graining averages over
many independent cells and the {\textsc{pdf}} approaches a Gaussian
form that, in the limit $\ell \sim L \to \infty$, has a vanishing
variance and its average is given by the global value of the
correlation.

As a consequence one expects to find similar fluctuations by taking a
system with finite size and examining the behavior of the run-to-run
fluctuations of the global two-time quantity (\ref{global}).  Indeed,
if $L\gg \xi$ one falls in case {\it ii}. Instead, if $L$ is of the
order of $\xi$, case {\it i} is found and the ``mesoscopic
fluctuations'' should be non-trivial. This argument justifies the
comparison of our approach to the ones in \cite{Sergio2} and
\cite{Luca2}.

\subsection{Variant of the Fredrickson-Andersen model}

In this section we present the analysis of the dynamics of the
kinetically facilitated model in $d=3$ choosing $p=3$ and then in
$d=2$ for $p=2$.  All the $3d$ numerical data have been obtained using
a cubic lattice of linear size $L=128$ lattice spacings at temperature
$T=0.4$ and the $2d$ data using a square lattice with $L=1000$ at
temperature $T=0.6$.  Coarse-grained data have been obtained using a
cubic coarse-graining volume of linear size $\ell=9$ lattice spacings
for $d=3$ and a square volume with $\ell=20$ for $d=2$. For $d=3$,
given the very large size of the system we used only one run (there is
no statistic average).

\begin{figure}[ht]
\center{\includegraphics[scale=.68]{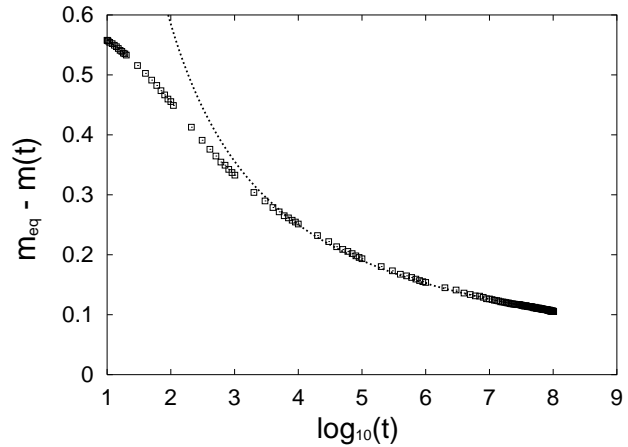}}
\caption{Time evolution of the magnetization density in the modified
  {\textsc{fa}} model. $d=3$, $L=128$, $T=0.4$.  The line is a fit to
  the curve given in Eq.~(\ref{magn}).}
\label{fig:magn}
\end{figure}

Let us first discuss the main qualitative features of the evolution of
the spin configuration before entering the quantitative analysis of
the time-dependent observables.  Initial configurations are random, in
that each spin points up or down with probability one half, which
mimics an infinitely rapid quench from $T=\infty$. Initially,
approximately $66\%$ for $d=3$ ($56\%$ for $d=2$) sites satisfy the
facilitation constraint and are able to flip with the heat bath
probability weight. At the early stages of the evolution it remains
relatively easy to find spins that satisfy the facilitation
constraint. As time evolves the number of positive spins, and hence
the magnetization, increases. However, as time passes one expects to
have isolated spins and small islands of spins that point
downwards. These may be remnants of the initial configuration or they
may be due to ``equilibrium-like'' thermal fluctuations. The former
are blocked until other down-pointing spins diffuse. The interior of
the latter can flip but their border might be blocked due to the
facilitated rule. A crossover (in the total time) between the ``easy''
initial regime and the ``hard'' latter regime was predicted
in~\cite{Graham}.

The analysis of the evolution for $d=3$ towards its equilibrium value
gives a first indication of the apparent ``glass transition''
temperature.  Figure~\ref{fig:magn} shows the time dependence of the
magnetization density at $T=0.4$ on a log-linear scale. We see the
above mentioned crossover occurring at $t_{\textsc{co}} \sim 10^3
{\textsc{mcs}}$. The value of the magnetization at the longest time
reached in the simulation, $t=10^8$ {\textsc{mcs}}, is still 10$\%$
away from the equilibrium value, $m_{\textsc{eq}}(T=0.4)\simeq 0.848$,
showing that the system is still far from equilibrium.  Assuming that
the curve found for this range of times does indeed approach the
equilibrium value and no further crossover to a different dynamic law
occurs (a hypothesis that might not hold, see~\cite{Mauro} for a more
detailed analysis) we find that the second slower regime is
well-characterized by the law
\begin{eqnarray}
  && m(t) \sim m_{\textsc{eq}}(T=0.4) - A \, \log_{10}^{-a} t
\; ,
\label{magn}
\;\;\;\; \\
\nonumber
&&
  m_{\textsc{eq}}(T=0.4)\simeq 0.848, \;
  A= 1.37, \;\; a=1.22 \;,
\end{eqnarray}
demonstrating very slow dynamics even in the behavior of this one-time
quantity.  At higher temperatures one finds that the maximum
magnetization reached in the numerical time window gets closer to the
equilibrium value and, for example at $T\sim 0.7$ it actually reaches
this value in $t\sim 10^7$ time steps~\cite{Mauro}. This value is
considerably lower than the putative glass transition identified by
Graham {\it et al.}~\cite{Graham}. These authors might have
overestimated this temperature by simulating very small systems (of
maximum size $V=16^3$) as compared to the ones we use ($V=128^3$).

\begin{figure}[ht]
\center{\includegraphics[scale=.68]{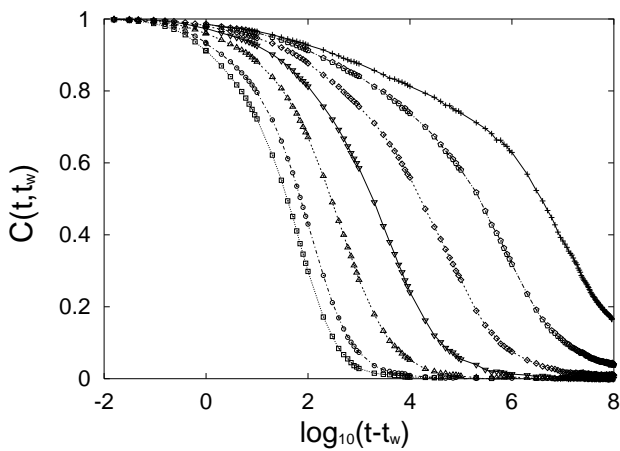}}
\caption{The global two-time normalized connected correlation
  function, $C(t,t_w)$, as a function of the logarithm of the time
  difference $t-t_w$, after a sudden quench to $T=0.4$. The curves
  correspond, from left to right, to seven logarithmically spaced
  values of the waiting-time, $t_w=10^1$, $10^2$, $10^3$, $10^4$,
  $10^5$, $10^6$, $10^7$ time steps. $d=3$, $L=128$.}
\label{fig:c_T04}
\label{fig:corr}
\end{figure}

What can we say about the structure of the remaining negative spins in
the sample at long times? A set of snapshots of the instantaneous
configuration shows that at around the crossover time there are no
finite size domains left but only isolated spins pointing downward or,
at most, very short strings of negative spins. It is then clear that
their annihilation needs the cooperative diffusion of other spins in
their neighbourhood, which is a very slow process indeed.

The aging curves for the global connected and normalized correlation
function defined in Eq.~(\ref{local-corr}) are shown in
Fig.~\ref{fig:c_T04} for several values of $t_w$ and as a function of
$t-t_w$ in log-linear scale. As discussed in Sect.~\ref{sect2},
disconnected correlation functions in glassy models usually show a
two-step relaxation with a first rapid decay towards a plateau, that
one estimates to be at $q_{\textsc{ea}}(T) \sim m_{\textsc{eq}}^2(T)$
for a model undergoing simple domain growth, and a second slower aging
decay towards zero. Connected correlation functions do not necessarily
show a plateau. This is the case in domain growth. For the times
explored numerically in the spin-facilitated model we do not see the
development of the plateau (which is not even seen on a log-log
scale). This might be due to the fact that the system is still very
far from an asymptotic dynamic regime or it may simply mean that the
connected correlation does not have a fast regime~\cite{Mauro}.

\begin{figure}[ht]
\center{\includegraphics[scale=.68]{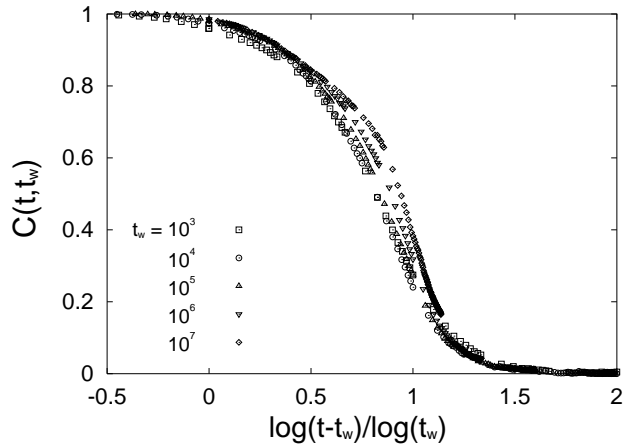}}
\caption{Attempt to scale the global connected and normalized
  correlation function presented in Fig.~\ref{fig:c_T04}. See the text
  for a discussion.}
\label{fig:c_T04_scaling}
\end{figure}

We have attempted to scale the correlation data for all time
differences and waiting times that are equal or longer than $t_w=10^3$
time steps (note that the curves for the two shorter waiting times are
clearly different from the rest). We found that the scaling
\begin{equation}
  C(t,t_w) \sim f \left( \frac{\ln(t-t_w)}{\ln t_w} \right)
\label{ultram}
\end{equation}
describes the data quite precisely for intermediate waiting times
($t_w=10^2$--$10^4$ time steps), but fails to collapse the data for
longer waiting times, as shown in
Fig.~\ref{fig:c_T04_scaling}. Interestingly, this scaling form is a
precursor of dynamic ultrametricity~\cite{Cuku2,Bertin}.  We have
searched for other simple scaling forms without finding any that
describe the full data in a satisfactory way. For this reason, we
shall use the value of the global correlation itself to scale the
{\textsc{pdf}}'s of the local fluctuations. The same difficulty arises
in the other kinetically constrained models that we have
studied. Henceforth we omit the subindex ${\textsc{s}}$ that indicates
the slow regime, since the kinetic facilitated models we have analyzed
do not seem to have a first fast regime in their connected
correlations.

\begin{figure}[ht]
\center{\includegraphics[width=\columnwidth]{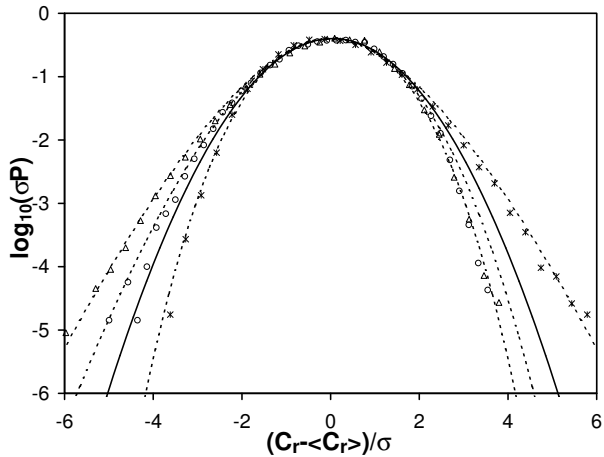}}
\caption{Normalized {\textsc{pdf}} of the local correlations
  $C_r(t,t_w)$ in the modified {\textsc{fa}} model, $d=3$, $L=100$,
  $\ell=5$ and $T=0.4$.  All curves drawn with symbols correspond to
  the waiting time $t_w=10^4$ {\textsc{mcs}} at various the time
  differences $t-t_w$.  The solid line represents the Gaussian normal
  form. The dashed lines are fits to modified Gumbel forms with
  parameter $a$ given by $a=13$ for $t-t_w=1$, $a=50$ for $t=1600$,
  and $a=-13$ for $t=400000$.}
\label{fig:pdf_T03tw1000M3}
\end{figure}

Let us now turn to the study of the fluctuations of the local
correlation functions, $C^o_r(t,t_w)$, as defined in
Eq.~(\ref{local-corr0}) and normalized to unity at equal times.  What
we clearly observe is that, even in this preasymptotic regime (in the
sense that the magnetization, a one-time quantity, is far from its
asymptotic value), the scaling of the {\textsc{pdf}} of local
correlations is the same as the scaling of the global
function. Moreover, the qualitative behavior of the form of the
{\textsc{pdf}}'s, their skewness and other momenta, are similar to the
ones that Cipelletti {\it et al.} found
experimentally~\cite{Luca2}. Working at different temperatures
or using different linear sizes does not modify this conclusion.

The summary of the features shown by the {\textsc{pdf}}'s of local
connected functions is the following.
Figure~\ref{fig:pdf_T03tw1000M3} shows a first indication of the
evolution of the {\textsc{pdf}} of local connected functions
(\ref{local-corr}) for fixed $t_w=10^4$ time steps and several values
of $t-t_w$. Initially, the curves are asymmetric with respect to their
average at short values of $t-t_w$, as exhibited by the curves for
$t-t_w=1$ and $10$ time steps.  When $t-t_w$ increases they slowly
approach a Gaussian normal form as shown by the curve for $t-t_w=1584$
time steps. Still later they go beyond the Gaussian form, and they
skew in the opposite direction, {\it e.g.} $t-t_w=63095$ time
steps. Finally, the curves very slowly tend to approach the Gaussian
form again, {\it e.g.} $t-t_w=398107$ time steps.

The skewed {\textsc{pdf}}'s resemble generalised Gumbel's distribution
(see App.~\ref{gumbel}). Even if this result is most probably just an
approximation, and we cannot give it a probabilistic interpretation,
it is worth mentioning the resemblance here. In Sect.~\ref{analytic}
we argue that these forms, and their evolution in time, can be
understood using an effective random manifold theory that naturally
gives rise to {\textsc{pdf}}'s with a Gumbel-like form.

The next point we investigate is whether the {\textsc{pdf}}'s of local
correlations in the slow regime of widely separated times scale as the
global correlation itself, namely
\begin{equation}
P(C_r;t,t_w) = g(C_r;C(t,t_w))
\; .
\label{eq:hypothesis}
\end{equation}
In~\cite{Castilloetal,Castilloetal2} we studied the $3d$ {\textsc{ea}}
model whose slow global correlation can be described with
$h_{\textsc{s}}(t)=t$. Thus, we checked if
\begin{eqnarray}
  P(C_r;t,t_w) &=& g(C_r;C_{\textsc{s}}(t,t_w)) \nonumber \\ &=&
  g\left[C_r; f \left(
  \frac{h_{\textsc{s}}(t_w)}{h_{\textsc{s}}(t)}\right)\right] \; .
\end{eqnarray}
Since we have not found a simple scaling form for global correlations
in the kinetically constrained models, we test the hypothesis
(\ref{eq:hypothesis}) directly. To this end, we chose pairs of total
and waiting times $t$ and $t_w$ such that the global correlation takes
the same value, say ${\cal C}$,
\begin{equation}
  C(t^{(k)},t_{w}^{(k)}) = {\cal C} \;\;\;\;\; \mbox{for} \;\;
  k=1,\dots, n
\end{equation}
and we plot the {\textsc{pdf}}'s of local correlations
\begin{equation}
  P(C_r(t^{(k)},t_{w}^{(k)})) \;\;\;\;\; \mbox{for} \;\; k=1,\dots, n
  \;.
\end{equation}
The hypothesis (\ref{eq:hypothesis}) implies that for fixed value of
${\cal C}$ these {\textsc{pdf}}'s should collapse onto a master curve.

\begin{figure}[ht]
\begin{center}
{\includegraphics[width=\columnwidth]{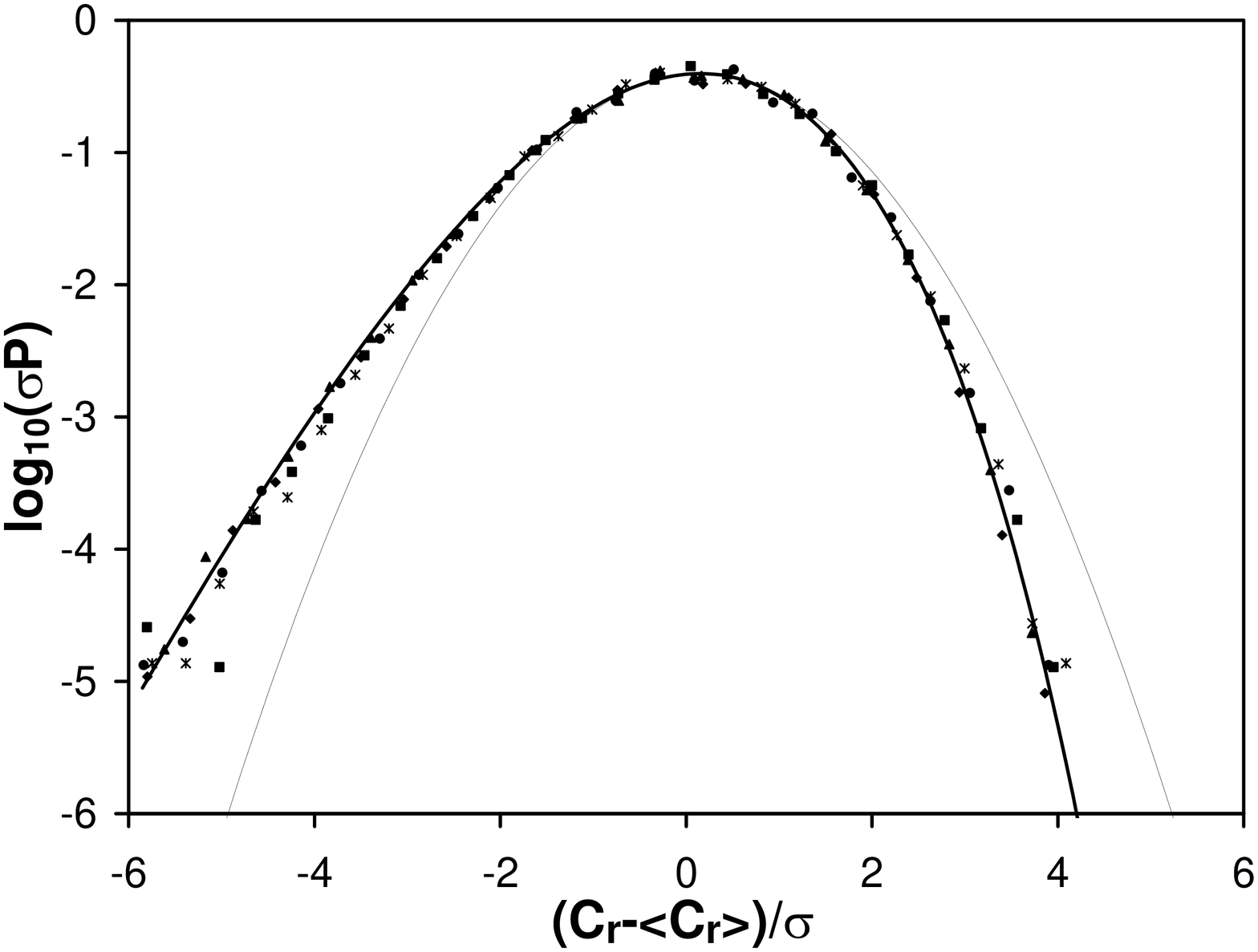}}\\
{\includegraphics[width=\columnwidth]{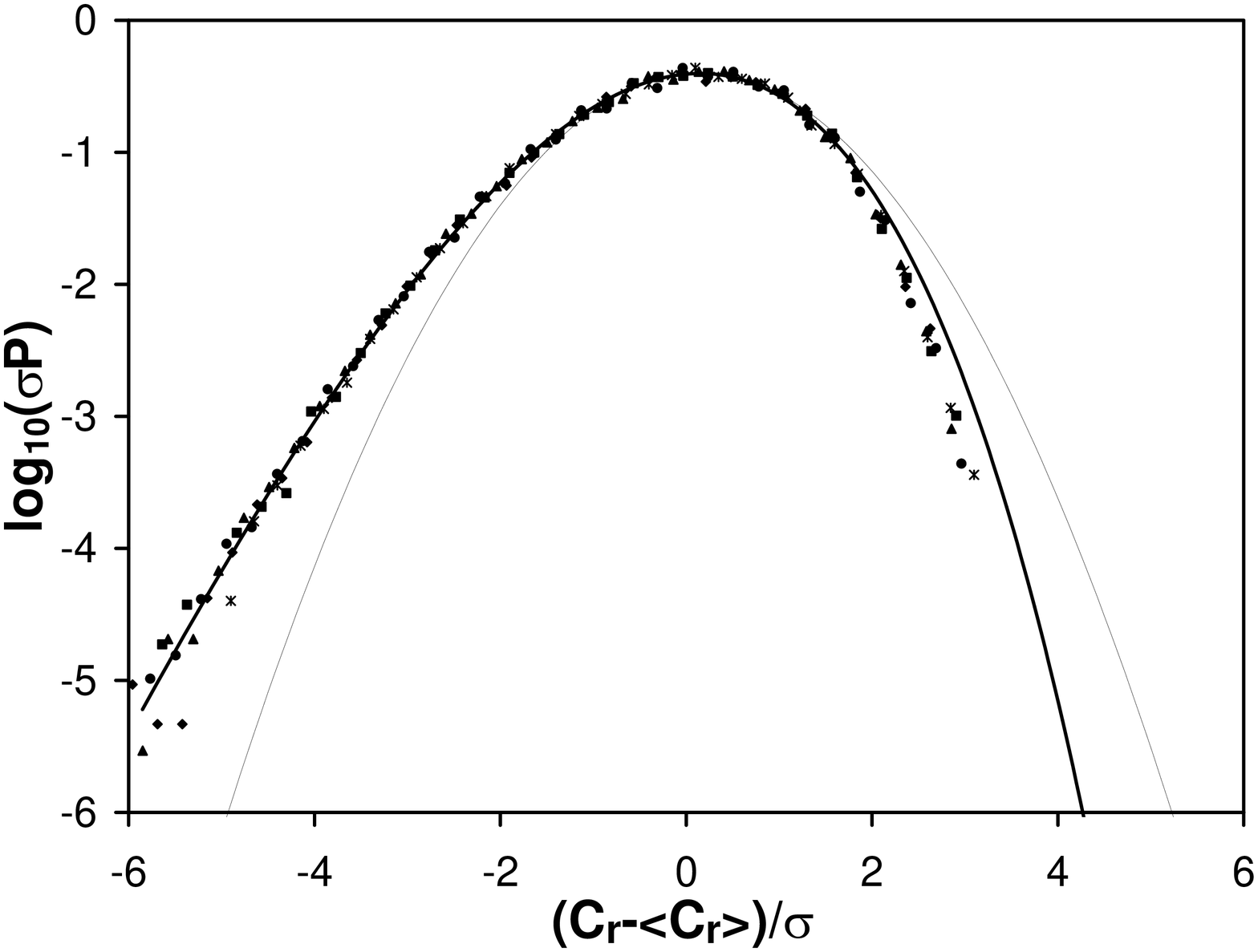}}
\caption{Scaling of the normalized {\textsc{pdf}} of local
    correlations in the modified {\textsc{fa}} model, $d=3$, $L=100$,
    $\ell=5$ and $T=0.4$.  ${\cal C}=0.88$ (top) and ${\cal C}=0.70$
    (bottom).  In both panels the data points correspond to five pairs
    of $t$ and $t_w$, with $t_w=10^1, \dots, 10^5$ time steps.  The
    thin line is a Gaussian normal form. The dark solid lines are fits to
    a Gumbel form with parameter $a=11$ (top) and $a=13$ (bottom).}
\label{fig:pdf_T04M4_scaling_C08-072}
\end{center}
\end{figure}
\begin{figure}[ht]
\begin{center}
{\includegraphics[width=\columnwidth]{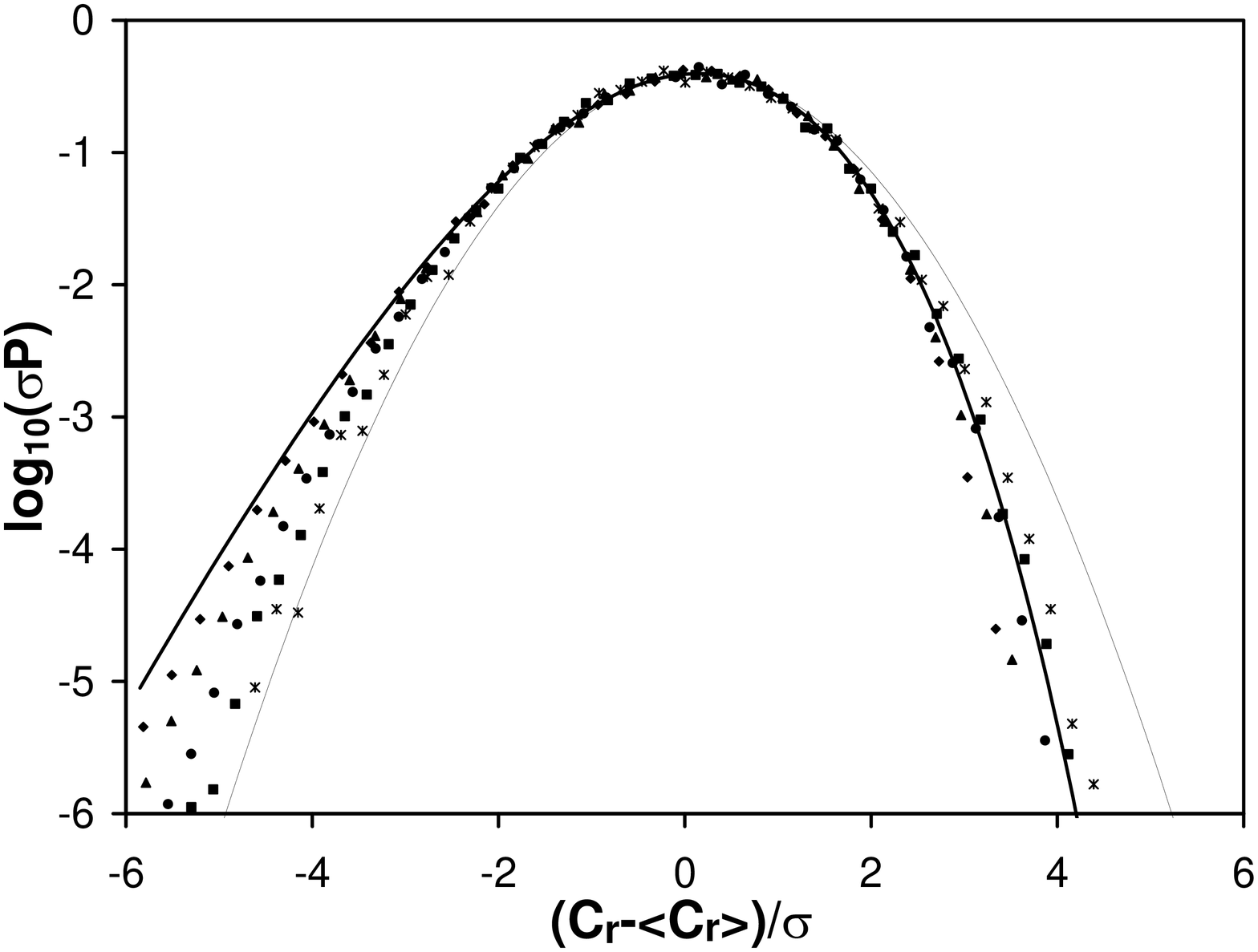}}\\
{\includegraphics[width=\columnwidth]{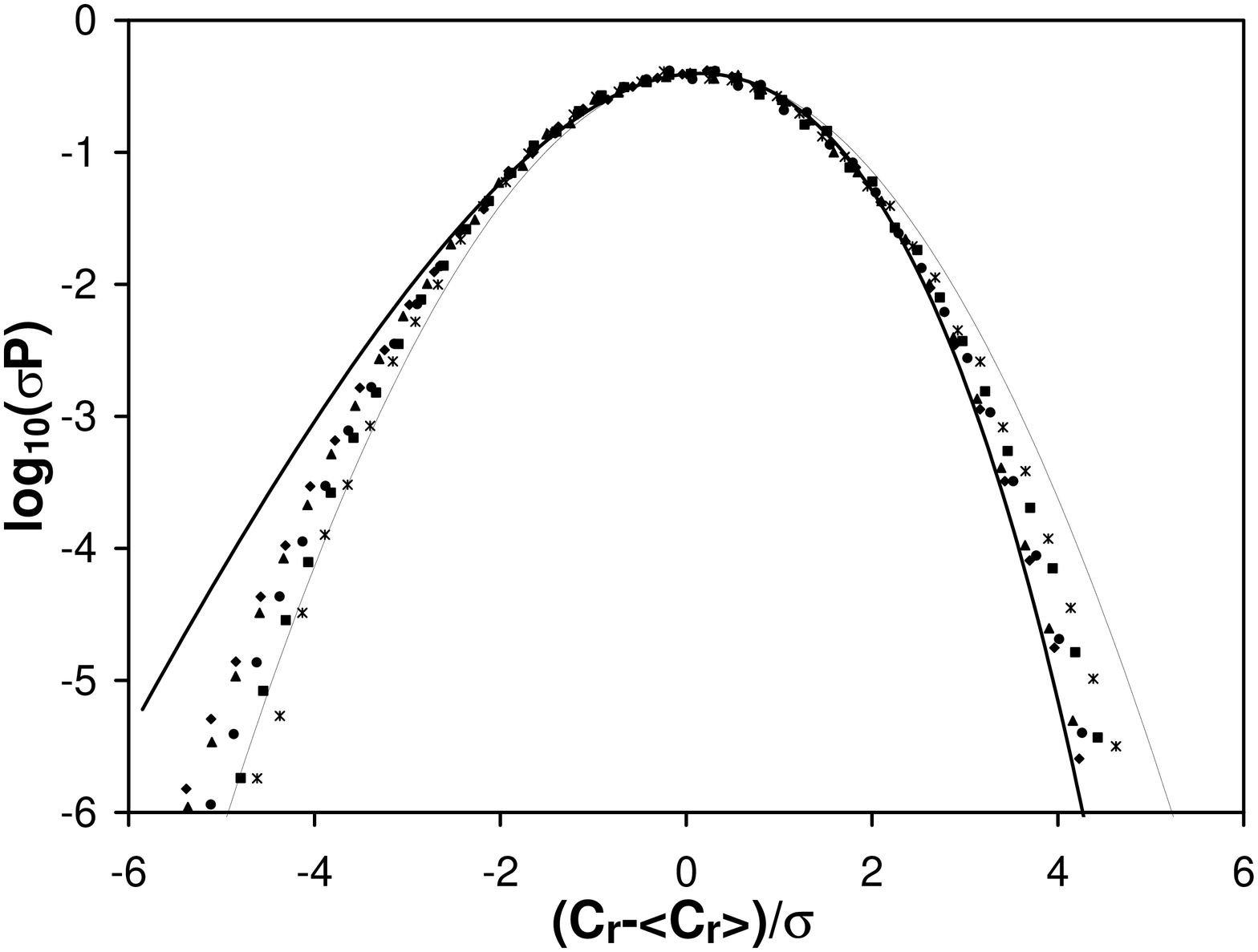}}
\caption{Scaling of the normalized {\textsc{pdf}} of local
  correlations calculated from the toy model for the parameters
  corresponding to the conditions described in
  Fig.~\ref{fig:pdf_T04M4_scaling_C08-072}. Notice that the lack of
  data collapse in the dynamics toy model shows that the collapse we
  find in Fig. 5 is a consequence of a non-trivial correlated
  dynamics.}
\label{fig:pdf_T04M4_scaling_C066-03}
\end{center}
\end{figure}

In Fig.~\ref{fig:pdf_T04M4_scaling_C08-072} we show the rather good
demonstration of master collapse onto the Gumbel form for the modified
FA model. To demonstrate that the strongly skewed distributions that
we observe arise from non-trivial dynamics induced by facilatation, we
construct a toy model that contains the same lattice features as the
FA model but evolves via unconstrained dynamics.  This model is
discussed in detail in Appendix B.  As shown in
Fig.~\ref{fig:pdf_T04M4_scaling_C066-03}, the distribution of local
correlations in this model is much weaker, and may be completely
attributed to binomial-like fluctuations due to small coarse-grained
cells.  Furthermore, the data show no regime of approximate collapse
onto a master curve, and for larger values of $t_{w}$ the
distributions rapidly evolve to a Gaussian form.  The non-trivial form
and approximate scaling of the PDF of local correlations results from
the dynamical constraints that induce a non-trivial dynamic length
scale.

\begin{figure}
\includegraphics[width=\columnwidth]{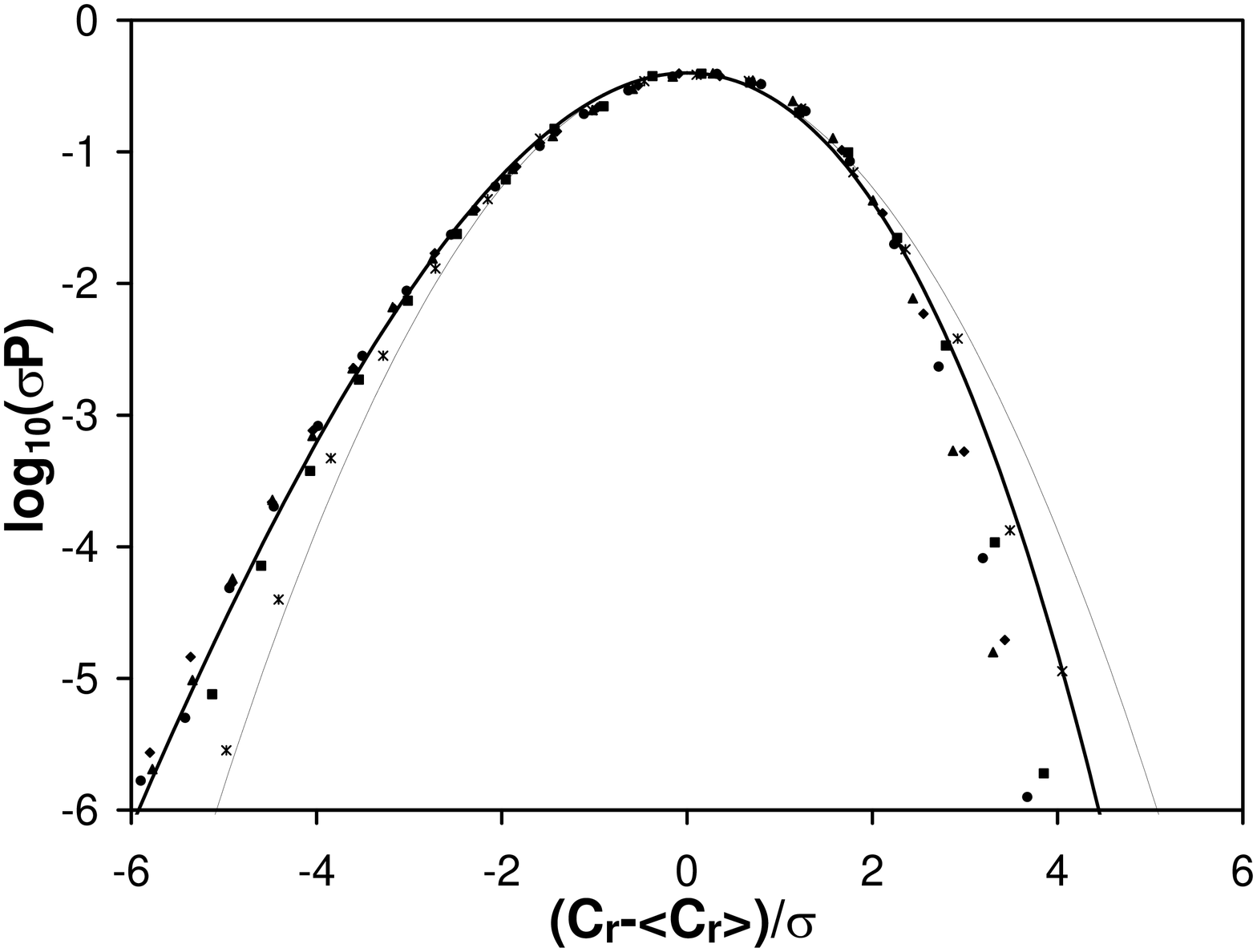}\\
\includegraphics[width=\columnwidth]{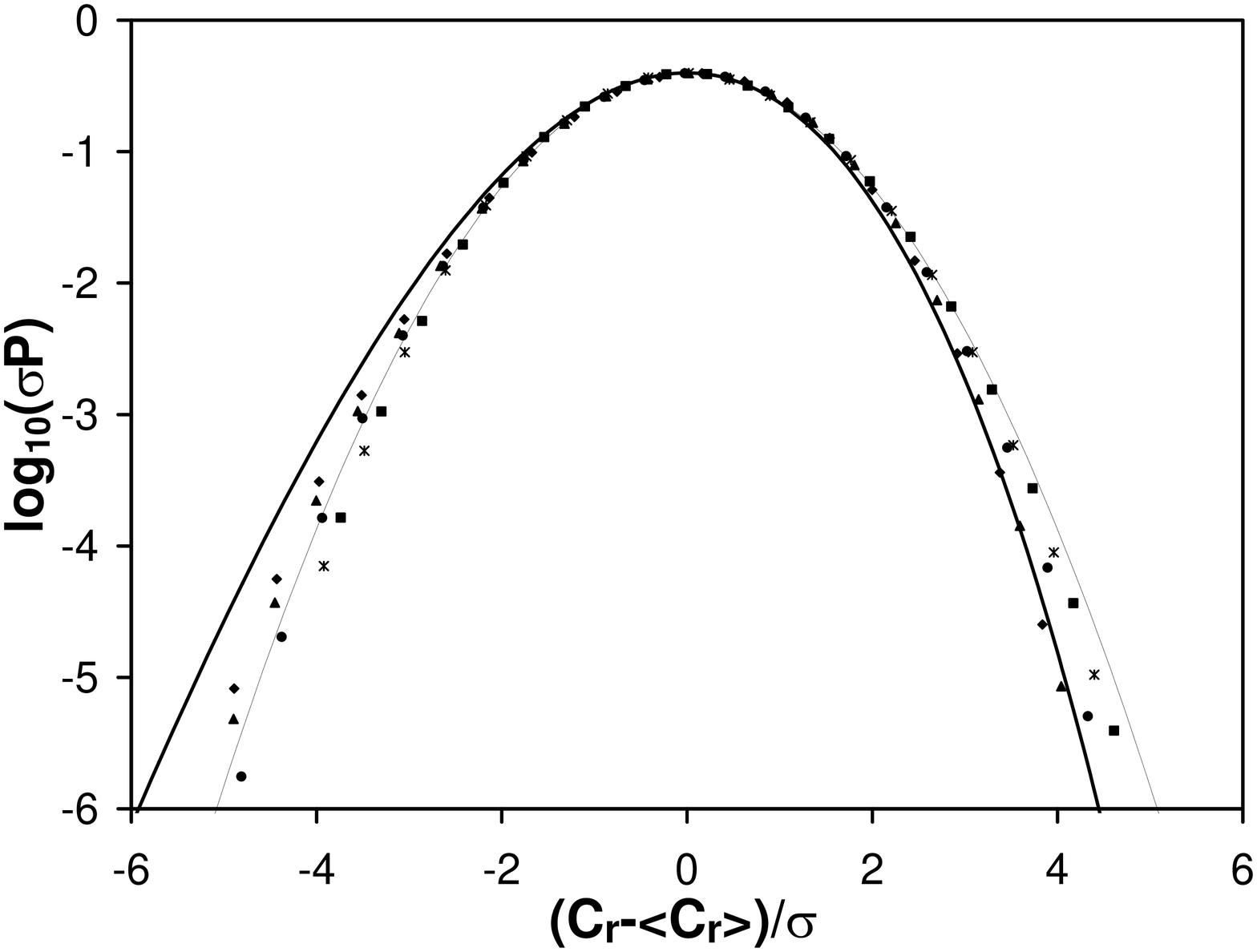}
\caption{Correlations in the modified {\textsc{fa}} model, $d=2$ and
  $L=1000$.  Top: Scaling of the
  {\textsc{pdf}}'s with ${\cal C}\sim 0.7$ at the same temperature for
  mesoscopic blocks of size $\ell=10$.  The data correspond to five
  pairs of $t$ and $t_w$, with $t_w=10^1,\ldots,10^5$ time steps. The
  dotted line is the Gaussian normal form.  The dark solid line is a fit to
  a Gumbel form with $a=35$. Bottom: Toy model results for conditions
  corresponding to the top panel.}
\label{fig:faccorrel}
\end{figure}

Now we look at the $d=2$, $p=2$ version of the model. In
Fig.~\ref{fig:faccorrel}, we show the behavior of the normalized
two-time connected correlation functions.  As in the case of $d=3$
$p=3$, we could find no global scaling of the decay for all waiting
times.  Figure~\ref{fig:faccorrel} also shows the collapse of local
correlations with ${\cal C}\sim 0.7$ onto the Gumbel form.
Interestingly, while the Gumbel parameters may depend sensitively on
temperature, dimensionality and the value of ${\cal C}$ for which the
collapse is plotted, the scaling onto the Gumbel form is remarkably
accurate.

\subsection{Edwards-Anderson disordered spin system}

\begin{figure}[ht]
\includegraphics[scale=.68]{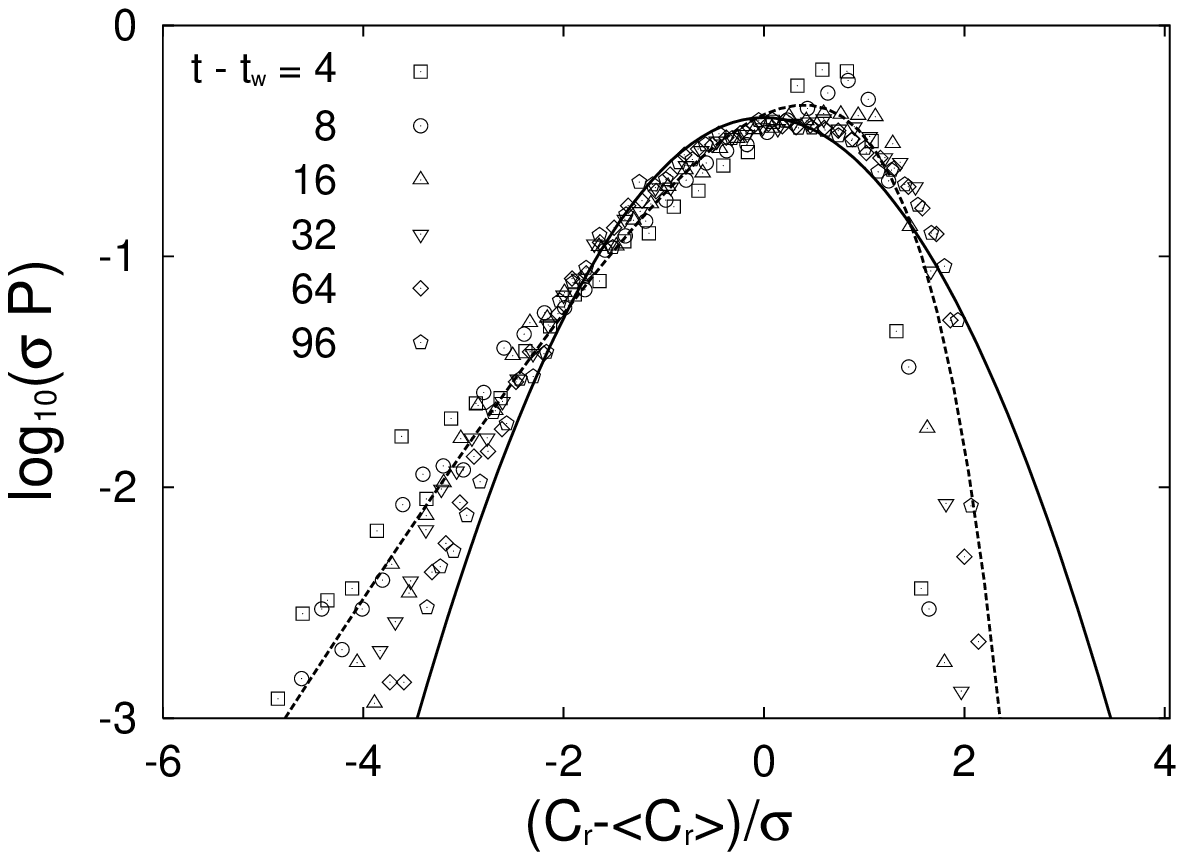}
\includegraphics[scale=.68]{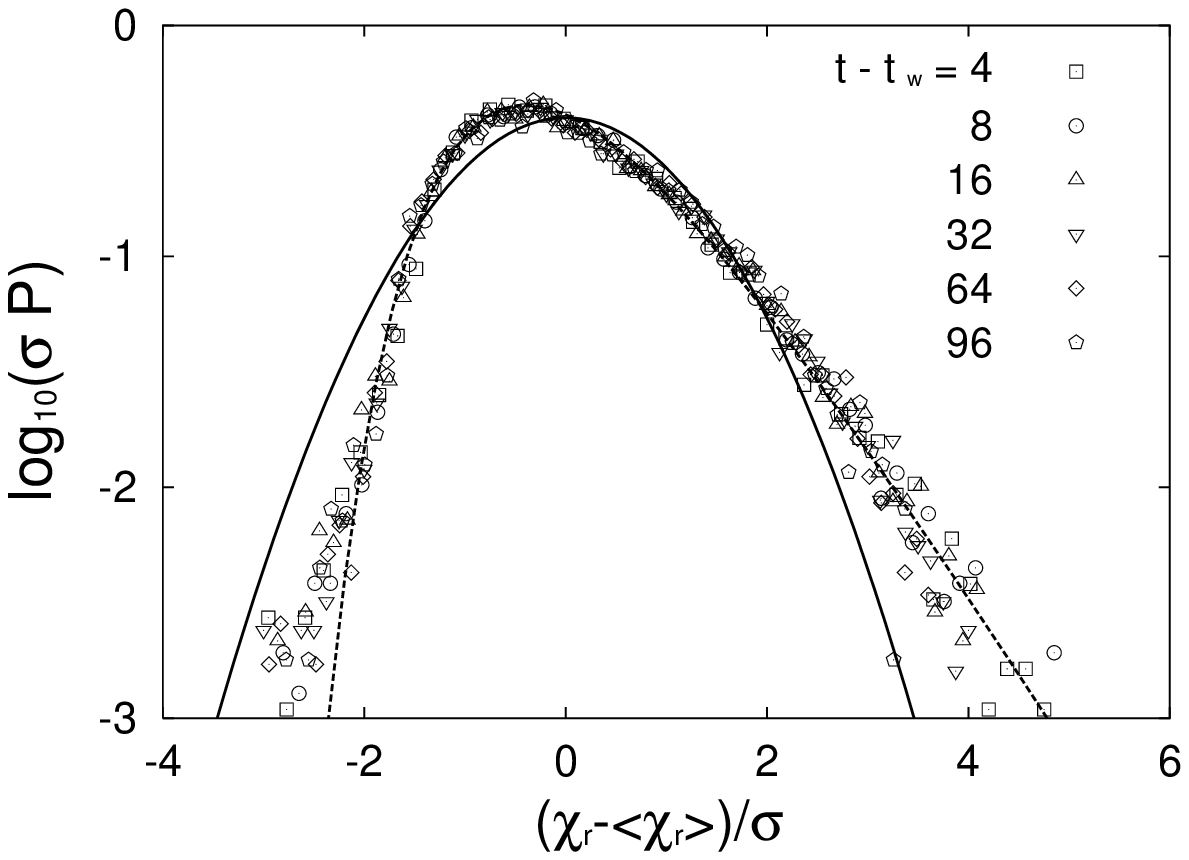}
\caption{Normalized {\textsc{pdf}}'s of global correlation $C_r$ and
  integrated response $\chi_r$ (as defined in~\cite{Castilloetal}) for
  a small $3d$ {\textsc{ea}} system with linear size $L=8$,
  temperature $T=0.4$ and waiting time $t_w=1000$.  Data averaged over
  $8000$ samples. The full line is the Gaussian distribution, while
  the dashed line is the generalised Gumbel form with $a=\pi/2$.}
\label{FIG:EA1}
\end{figure}

Similar results have been obtained for the disordered {\textsc{ea}}
spin-glass model with usual Monte Carlo dynamics. In
Fig.~\ref{FIG:EA1} we show the {\textsc{pdf}}'s of global correlations
and integrated responses for a small system.  In this case, since the
system is not macroscopic, we explore ``mesoscopic'' fluctuations that
include similar information as the local fluctuations when $L$ is not
much larger than the correlation length. We see that the generic
behavior of the data is very similar to the one displayed in the
previous subsection, although the particular time scaling [of the form
(\ref{global-C})-(\ref{slow-C}) in this case] is different. The data
is also described for time differences that are shorter than the
characteristic relaxation time by a negatively skewed form that
resembles a modified Gumbel distribution. The trend in $t-t_w$ is the
same one although the approach to the Gaussian form is much slower and
this asymptotic form is hardly reached within the time window
explored.  We have also simulated the procedure used by Cipelletti
{\it et al.}, {\it i.e.} we have measured temporal fluctuations in
mesoscopic samples, and found very similar results.

It should be emphasized again that the glassy physics behind the EA
model and the FA model differs in that a spin-glass transition exist
at finite temperatures in the EA model, while no thermodynamic
transition exists at finite temperature in the FA model.  The growing
correlation length in the FA model is of a dynamical nature.  We have
verified that the Gumbel-like distributions systematically revert to
Gaussian distributions when the cell size is made larger.  This
suggests a direct way to estimate dynamical correlation lengths.  Work
along these lines is underway and will be reported in a future
publication.

\section{Analytic arguments}
\label{analytic}

In this section we recall and extend some of the theoretical arguments
introduced in \cite{Castilloetal,Castilloetal2,Chamonetal} to describe
local and mesoscopic fluctuations in glassy systems. In those articles
we focused on the time scaling of the {\textsc{pdf}}'s without paying
special attention to the form of these distributions. Here, instead,
we investigate their form and we develop an effective theory to
describe them.

\subsection{Reparametrization invariance}

The two-time scaling of correlation functions that are monotonic
functions of the two times take the form
\begin{equation}
  C(t,t_w) \sim f\left( \frac{k(t_w)}{k(t)}\right)
\label{generic-scaling}
\end{equation}
within each ``correlation scale'' as defined in~\cite{Cuku2}.  In
Sect.~\ref{fluctuations} we discussed the dynamic evolution of three
kinetically constrained models and the $3d$ {\textsc{ea}}
spin-glass. For the latter, the aging decay is described quite well by
a two-scale relaxation, see Eqs.~(\ref{global-C}) and (\ref{slow-C}),
with a ``simple aging'' form, $h_{\textsc{s}}(t)=t$, for the slow
part. For the kinetically constrained models we have not found a
single function $h_{\textsc{s}}(t)$ that describes the full slow decay
from ${\cal C}=1$ to ${\cal C}=0$ for all waiting times and times
explored.

In order to keep the presentation as simple as possible, we discuss
the theoretical arguments that allow one to predict the behavior of
local fluctuations, assuming that the full slow decay is represented
by the scaling given in (\ref{generic-scaling}) with a single function
$k(t)=h_{\textsc{s}}(t)$. By the end of this section we argue how this
approach can be extended to describe the case of a more complicated
slow relaxation that does not fall into this category.  (The failure
may be due to the development of many correlation scales in the slow
relaxation.)

In~\cite{Chamonetal} it has been shown that the dynamic action
controlling the local coarse-grained dynamics at long waiting time
$t_w$ and long time differences $t-t_w$ in the {\textsc{ea}} model
approaches a form that is invariant under {\it global}
reparametrizations of time that act as
\begin{equation}
  t \to h(t) \; ,
\end{equation}
with $h(t)$ a monotonically growing function. This change transforms
the ``fields'' in this action in a special way that, at the level of
expectation values, corresponds to leaving invariant the local
correlations, $C_r(t,t_w) \to C_r(h(t),h(t_w))$, and the integrated
linear responses, $\chi_r(t,t_w) \to$$ \chi_r(h(t),h(t_w))$.  Symmetry
breaking terms exist but they become smaller and smaller (irrelevant)
as times and time differences increase. These symmetry breaking terms
(which have their origin in the details of the short-time and
short-time-difference dynamics), although scaling down to zero at long
times, are responsible for selecting a particular parametrization for
which the global ${\cal C}$ decays from $q_{\textsc{ea}}$ to 0 as $t$
is increased. Notice that if reparametrization invariance is left {\it
unbroken}, ${\cal C}$ must be a constant independent of the
times. Without entering the details of this action, let us introduce
the main consequences of the existence of this approximate invariance.

Due to the global invariance, and the scaling in
(\ref{generic-scaling}), one can argue that the slow part of the
coarse-grained {\it local} correlations scale in time
as~\cite{Castilloetal,Castilloetal2,Chamonetal}:
\begin{equation}
  C^{\textsc{s}}_r(t,t_w) \sim
  f\left( \frac{h_r(t_w)}{h_r(t)}\right) \;,
\label{scaling1}
\end{equation}
where the functions $h_r(t)$ are {\it local} time reparametrizations,
$t \to h_r(t)$.  The superscript ${\textsc{s}}$ means that we are
looking here at the slow part of the decay of the local correlation
only (in other words, the decay for long time differences compared to
the waiting time $t_w$). The argument in~(\ref{scaling1}) varies from
unity when the times tend to be equal, $t\to t_w^+$, to zero at widely
separated times $t\gg t_w$. The external function $f$ takes values
between $f(1)=q_{\textsc{ea}}$ and $f(0)=0$.  The first limit is
independent of the position $r$ since one does not expect the plateau
to be a fluctuating quantity if a sufficiently large coarse-graining
volume is used.  The second limit holds if there is no external field
applied and all local correlations decay to zero. We consider here
monotonic local reparametrizations of the times, such that $h_r(t_w)<
h_r(t)$ at all $r$ for $t_w<t$.  This is equivalent to assuming that
the time lags we shall consider are not extremely short.

Our original claim was that the external function $f$ is the same for
all coarse-grained centers in the sample, while all spatial
fluctuations are encoded in the internal function $h_r$.  The reason
for this proposal is that the {\it global} reparametrization
invariance in time of the dynamic action in this two-time regime leads
to low action excitations (Goldstone modes) for smoothly varying {\it
  spatial} fluctuations in the reparametrization of time, but not in
the external form of the correlations. As in a sigma model, the
external function $f$ fixes the manifold of states, and the local time
reparametrizations correspond to fluctuations restricted to this fixed
manifold of states. The relation to the sigma model is described in
detail in Ref.~\cite{Castilloetal2} and we shall not repeat it here.

The symmetry-based argument and its predictions, originally developed
on the basis of the time-reparametrization invariance of the dynamical
action of the {\textsc{ea}} model, was proposed to hold beyond this
model whenever a separation between fast and slow decay in the global
correlation and response develops in time. Our results on models
without quenched disorder conclusively demonstrate the generality
beyond the {\textsc{ea}} model.

\subsection{Effective random manifold action}

As in the sigma model approach to an interacting system in which we
have identified a relevant variable and a relevant symmetry, we
propose an effective action for the relevant degree of freedom that
quantifies the fluctuations about the globally symmetric result.  The
philosophy is the same as the one followed when describing spin-wave
excitations in a Heisenberg ferromagnet with a quadratic action that
depends on the angular variables only.

The scaling form in Eq.~(\ref{scaling1}) can be written in an
equivalent form by defining $\phi_r(t) \equiv \ln h_r(t)$:
\begin{eqnarray}
  C^{\textsc{s}}_r(t,t_w) &\sim& f\left( {\rm
e}^{-[\phi_r(t)-\phi_r(t_w)]}\right)\nonumber\\ &=&f\left( {\rm
e}^{-\int_{t_w}^t dt' \dot\phi_r(t')}\right)\nonumber\\ &\equiv&
f\left({\rm e}^{- \Delta {\phi_r} |_{t_w}^t } \right) \; .
\label{local-corr-param}
\end{eqnarray}
In the dynamic case under study, after parametrizing the correlation
in this way, the relevant field is the function $\phi_r(t)$.  The
action governing the dynamics of $\phi_r(t)$ depends on the details of
the particular problem. However, we can greatly restrict the form of
the possible actions by simply using the constraints due to the
symmetries. These are:

\begin{enumerate}

\item[{\it i.}] The action must be invariant under a global
time reparametrization $t\to s(t)$.

\item[{\it ii.}] If our interest is in short-ranged problems, the
action must be written using local terms. The action can thus contain
products evaluated at a single time and point in space of terms such
as $\phi_r(t)$, $\dot\phi_r(t)$, $\nabla\phi_r(t)$,
$\nabla\dot\phi_r(t)$, and similar derivatives.

\item[{\it iii.}] The scaling form in Eq.~(\ref{local-corr-param}) is
invariant under $\phi_r(t)\to\phi_r(t)+\Phi_r$, with $\Phi_r$
independent of time.  Thus, the action must also contain this
symmetry.

\item[{\it iv.}] The action must be positive definite.

\end{enumerate}

These requirements largely restrict the possible actions. The action
with the smallest number of spatial derivatives (most relevant terms)
is

\begin{equation}
S_{\textsc{eff}}=\int d^dr \int dt \left[
K\; \frac{\left(\nabla\dot\phi_r(t)\right)^2}{\dot\phi_r(t)}+ M\; \dot\phi_r(t)
\right]\; .
\end{equation}
The last term is simply a total derivative, and its space-time
integral is constant, so we shall drop it. Hence, we find that the
strong constraints imposed on the action reduces it to the form
\begin{equation}
S_{\textsc{eff}}=K\; \int d^dr \int dt\;
\frac{\left(\nabla\dot\phi_r(t)\right)^2}{\dot\phi_r(t)}
\; .
\label{S-effective}
\end{equation}
Notice that the action now solely depends on the time derivatives
$\dot\phi_r(t)$.

Let us now examine the consequences of this action on the form of the
local correlations (\ref{local-corr-param}). Due to the simple form
(\ref{S-effective}) the $\dot\phi_r$ are uncorrelated at any two
different times $t_1$ and $t_2$. Thus the expression
$\Delta\phi_r|_{t_w}^t=\int_{t_w}^t dt' \dot\phi_r(t')$ entering the
exponential in the scaling form in Eq.~(\ref{local-corr-param}) is a
sum of uncorrelated random variables in time. Hence, one can interpret
such expression as the displacement of a random walker with position
dependent velocities. Alternatively, one can think of the
space-dependent differences $\Delta\phi_r|_{t_w}^t=\int_{t_w}^t dt'
\dot\phi_r(t')$ as the net space-dependent height (labeled by $t$) of
a stack of spatially fluctuating layers $dt' \dot\phi_r(t')$. The
action for the fluctuating surfaces of each layer is given by
Eq.~(\ref{S-effective}).

Before proceeding, let us take a moment to show that indeed the action
in Eq.~(\ref{S-effective}) satisfies all the four constraints
enumerated above. First, consider a global change of variables $t\to
s(t)$, for which the action transforms as
\begin{eqnarray}
  S_{\textsc{eff}}&\to& K\; \int d^dr \int dt\;
  \frac{\left(\frac{ds}{dt}\right)^2\;
  \left(\nabla\dot\phi_r(s(t))\right)^2}{\left(\frac{ds}{dt}\right)\;
  \dot\phi_r(s(t))} \nonumber \\
  &=& K\; \int d^dr \int ds\; \frac{\left(\nabla\dot\phi_r(s)\right)^2}{\dot\phi_r(s)}
  = S_{\textsc{eff}}\;.
\end{eqnarray}
Hence, as we required in point {\it i.}, the action is invariant under
the global time reparametrization. The action is also clearly local.
Because it only contains explicitly $\dot\phi_r$ but not $\phi_r$, it
is invariant under $\phi_r(t)\to\phi_r(t)+\Phi_r$, with $\Phi_r$
independent of time. Finally, since the local reparametrizations
$h_r(t)$ are monotonically increasing functions of time, so are
$\phi_r(t)=\ln h_r(t)$, and consequently $\dot\phi_r(t)>0$. Hence, the
combination $(\nabla\dot\phi_r)^2/\dot\phi_r$ is always positive, and
the action is positive definite.

To better understand the consequences of this effective action, we
study it in two regimes. The first is one in which we consider only
small spatial fluctuations of the local reparametrizations with low
velocities. In the second, we consider the full effect of the action,
without assuming that the local fluctuations are small. We will use
the latter to obtain the full form of the probability distribution of
local correlations and compare it to the numerical and experimental
results.

\subsubsection{Linear regime}
\label{sec:linear}

We begin by making some simplifying assumptions so as to obtain a
better understanding of the consequences of local
reparametrizations~\cite{Castilloetal,Chamonetal}. We shall shortly
relax these assumptions. If we assume that the site-to-site
fluctuations are small, the field $\phi_r(t)$ will be given by
\begin{equation}
  \phi_r(t) = s(t) + \varphi_r(t) \;\;\; \mbox{with}
  \;\;\; \varphi_r(t) \ll s(t) \; ,
\label{eq:approx1}
\end{equation}
and the uniform part $s(t)$ defined as
\begin{equation}
  s(t)\equiv \ln h_{\textsc{s}}(t)
\end{equation}
with $h_{\textsc{s}}(t)$ the scaling function of the global correlation.
In terms of the $s$ parametrization we can write
\begin{equation}
  \phi_r(s) = s + \varphi_r(s) \;,
\end{equation}
and
\begin{equation}
  \dot\phi_r(s) = 1 + \dot\varphi_r(s) \;.
\label{linear-approx}
\end{equation}
We shall further assume that
\begin{equation}
  \;\;\; \dot\varphi_r(s) \ll 1 \; .
\label{linear-approx1}
\end{equation}
This approximation leads to the action
\begin{equation}
  S_{\textsc{eff}}\approx K\; \int d^dr \int ds\;
  \left(\nabla\dot\varphi_r(s)\right)^2 \;.
\label{S-effective-linear}
\end{equation}
The surfaces $\dot\varphi_r(s)$ are uncorrelated for different
reparametrized times $s$. The argument that enters in the evaluation
of the local correlation functions $C_r(t,t_w)$ is, according to
Eq.~(\ref{local-corr-param}),
\begin{eqnarray}
  \Delta\phi_r|^t_{t_w}&=&\int_{s(t_w)}^{s(t)} ds \left[1+
  \dot\varphi_r(s) \right] \nonumber\\
  &=&  s(t)-s(t_w)\;+\;\sqrt{s(t)-s(t_w)}\;X_r \nonumber\\
  &=&  \ln\frac{h_{\textsc{s}}(t)}{h_{\textsc{s}}(t_w)}\;
  +\;\sqrt{\ln\frac{h_{\textsc{s}}(t)}{h_{\textsc{s}}(t_w)}}\;X_r \;,
\label{linear-collapse}
\end{eqnarray}
where the square root arises from the fact that the $\dot\varphi$
surfaces are uncorrelated for different $s$, leading to a sum of
independent variables in the reparametrized time direction. The $X_r$
is a Gaussian random surface that encodes the spatial fluctuations,
and it is distributed with probability
\begin{equation}
  P(X_r)\propto {\rm e}^{-K\; \int d^dr\; (\nabla X_r)^2 } \;.
\label{eq:X_r-dist}
\end{equation}

Even within the linear approximation (\ref{linear-approx}) and
(\ref{linear-approx1}), one can already explain why the probability
distributions for local correlations of a system with a global scaling
as in (\ref{global-C})-(\ref{slow-C}) collapse as a function of
$h_{\textsc{s}}(t)/h_{\textsc{s}}(t_w)$. This fact is a consequence of
Eqs.~(\ref{local-corr-param}) and
(\ref{linear-collapse})~\cite{Castilloetal,Castilloetal2}.  Moreover,
noticing that the two-time global correlation is in one-to-one
correspondence with the ratio $h_{\textsc{s}}(t)/h_{\textsc{s}}(t_w)$,
one concludes that the {\textsc{pdf}} of local correlations scales in
time as the global correlation itself.

While this approximation gives the observed time scaling of the
{\textsc{pdf}}'s of local correlators, it does not give their correct
shape, in particular when $t\to t_w^+$. The reason is that the
assumption that site-to-site fluctuations and their velocities are
small as compared to the bulk average fails more severely in this
regime. As we show next, we can still tackle the problem in the
non-linear regime.

\subsubsection{Non-approximate treatment of fluctuations}

In order to get a more accurate prediction of the form of the
{\textsc{pdf}}, one needs to go beyond the approximations in
Eqs.~(\ref{eq:approx1}) and (\ref{linear-approx1}).  As done in the
previous section, we start by writing the action in the
parametrization $s(t)=\ln h_{\textsc{s}}(t)$ chosen by the bulk
values, and by defining a new field~\cite{footnote},
%
%
$\psi_r(s)$, such that $\psi^2_r(s)={\dot\phi_r(s)}$, one then has
\begin{equation}
  S_{\textsc{eff}}=K\; \int d^dr \int ds\;
  \left(\nabla\psi_r(s)\right)^2 \;.
\label{S-effective-psi}
\end{equation}
This is again the action of uncorrelated Gaussian surfaces for
different proper times $s$.

The argument entering in Eq.~(\ref{local-corr-param}) now reads
\begin{eqnarray}
  \Delta\phi_r|^t_{t_w}&=&\int_{s(t_w)}^{s(t)} ds\; \dot\phi_r(s)
  =\int_{s(t_w)}^{s(t)} ds \;\psi^2_r(s) \; ,
\label{nonlinear-collapse}
\end{eqnarray}
Due to the Gaussian statistics of the $\psi_r(s)$, it is simple to
show that connected $N$-point correlations of
$\Delta\phi_{r_1}|^t_{t_w}$ satisfy
\begin{eqnarray}
  \langle \Delta\phi_{r_1}|^t_{t_w}\; \Delta\phi_{r_2}|^t_{t_w} \cdots
  && \Delta\phi_{r_N}|^t_{t_w} \rangle_c
  = \\\nonumber &&[s(t)-s(t_w)]\; {\cal F}(r_1,r_2,\dots,r_N) \; ,
\label{N-corr}
\end{eqnarray}
where the function ${\cal F}$ can be obtained from Wick's theorem,
summing over all graphs that visit all sites (connected) with two
lines (because of $\psi^2$) for each vertex $i$ corresponding to a
position $r_i$. Notice that the reparametrized times appear only in
the prefactor $\Delta s=s(t)-s(t_w)$. Therefore time dependencies are
functions of $\Delta s$ alone. Once again we find that the
probabilistic features of the fluctuations of local correlations are
functions of the scaling variable, $\Delta s=\ln
h_{\textsc{s}}(t)/h_{\textsc{s}}(t_w)$, and hence of the global
correlation itself.  This coincides with the numerical observations.

Here, a comment is in order. In obtaining the effective action, we
have neglected higher gradients and time derivatives because these
were considered irrelevant. In particular, this is justified for large
reparametrized time differences. Note, however, that these terms are
important when $\Delta s$ becomes of order unity.  This short
(reparametrized) time-cutoff scale means that the surfaces for
different $s_{1,2}$ are uncorrelated only when $|s_1-s_2|$ is larger
than a cutoff $\delta s$. Keeping this in mind, let
\begin{equation}
  \Delta\phi_r|^t_{t_w}=\delta s\;\sum_{i=1}^n\; \psi^2_r(s_i) \;,
\label{discrete-nonlinear-collapse}
\end{equation}
where we have discretized the integral in
Eq.~(\ref{discrete-nonlinear-collapse}) in steps of the cutoff $\delta
s$. The number of such steps is $n=\Delta s/\delta s$. The expression
(\ref{nonlinear-collapse}) and the scaling relation
(\ref{local-corr-param}) allow us to compute the {\textsc{pdf}} for
the local correlation functions (these expressions can also be used to
investigate $N$-point correlations of $C_r(t,t_w)$). We shall do this
in the next section generating numerically random Gaussian surfaces
$\psi_r(s)$, from which we obtain $\Delta\phi_r|^t_{t_w} $ and then
$C_r(t,t_w)$.  Let us now discuss the extreme cases (when $n$ is
either small or large) before turning to the numerical calculations,
which are better suited to describe the crossover between these two
limits.

\vspace{0.5cm}
\noindent
\underline{Case ${\cal C}\approx q_{\textsc{ea}}$}
\vspace{0.5cm}

If the global correlation is close to the Edwards-Anderson parameter,
$n=\Delta s/\delta s$ is small.  Consider for simplicity $n=1$. In
this case, letting $Y_r= \psi_r(s_1)$, we write
\begin{equation}
  \Delta\phi_r|^t_{t_w}={\delta s}\; Y^2_r \;,
\end{equation}
with $Y_r$ distributed according to
\begin{equation}
  P(Y_r) \propto {\rm e}^{-K\;\int d^dr\; (\nabla Y_r)^2 } \;.
\label{eq:Y-dist}
\end{equation}
Using Eq.~(\ref{local-corr-param}) the local correlation is given by
\begin{equation}
C_r(t,t_w)=f\left({\rm e}^{-\Delta\phi_r|^t_{t_w}}\right)=
f\left({\rm e}^{-{\delta s}\; Y^2_r}\right)
\;.
\label{eq:C-func}
\end{equation}
In what follows, we connect the {\textsc{pdf}}'s of the local
correlations $C_r$ coarse grained over a volume around the point ${\bf
r}$ and the {\textsc{pdf}}'s for the fluctuations of the global
magnetization $m$ of the {\textsc{xy}} magnet at low temperature, as
studied by Bramwell {\it et al.}~\cite{Peter}, or the roughness of a
Gaussian surface of the Edwards-Wilkinson type, as analyzed by R\'acz
{\it et al.}~\cite{Racz}. For concreteness, we use in the following
the language of the {\textsc{xy}} model.

In the case of the {\textsc{xy}} magnet, the local field $\theta_r$ is
distributed according to
\begin{equation}
  P(\theta_r)\propto {\rm e}^{-K\;\int d^dr\; (\nabla \theta_r)^2} \;,
\label{eq:theta-dist}
\end{equation}
and one can write the local magnetization in one direction as
\begin{equation}
  m_r=\cos \theta_r \;,
\label{eq:bhp-mag}
\end{equation}
where for simplicity we took $m=1$.  Bramwell {\it et al.} found that
the distribution of the global magnetization of finite-size systems
does approach a scale-invariant non-Gaussian distribution (when the
magnetization is normalized by subtracting the average and dividing by
the standard deviation). In particular, the {\textsc{pdf}} is skewed
and can be approximated by a generalized Gumbel distribution.  The
underlying reason for the non-applicability of the central limit
theorem is that the system is critical (there is no finite correlation
length) and hence the local magnetizations are not uncorrelated.
Similarly, in a system where the correlation length $\xi$ is finite,
non-Gaussian distributions will be found as long as the coarse
graining length is smaller than $\xi$.

The connection between the {\textsc{pdf}}'s $P(C_r)$ and $P(m)$ for
dynamic local coarse grained correlations $C_r(t,t_w)$ in our case and
the global magnetization $m$ in the {\textsc{xy}} model is as follows.
First, the distribution of the variables $Y_r$ and $\theta_r$ in
Eqs.~(\ref{eq:Y-dist})~and~(\ref{eq:theta-dist}) are identical. The
fact that the functional dependencies of $C_r$ and $m_r$ on $Y_r$ and
$\theta_r$ respectively, as given in
Eqs.~(\ref{eq:C-func})~and~(\ref{eq:bhp-mag}), are not the same lead
to slightly different forms of the {\textsc{pdf}}'s that, in both
cases, can be approximated by generalized Gumbel forms with different
parameters. To illustrate this point, consider the case in which the
stiffness $K$ is large so that both variables $Y_r$ and $\theta_r$
(centered around zero) are small within a given coarse graining
volume. In this case, we can expand
Eqs.~(\ref{eq:C-func})~and~(\ref{eq:bhp-mag}) as follows:
\begin{eqnarray}
  C_r&=& f\left({\rm e}^{-{\delta s}\; Y^2_r}\right) \approx f(1) -
\delta s \;Y_r^2\; f'(1) +\dots
\label{eq:expansion-C}
\\
m_r&=& \cos \theta_r \approx 1 - \frac{1}{2}\; \theta_r^2 +\dots
\label{eq:expansion-m}
\end{eqnarray}
and because the {\textsc{pdf}}'s are calculated for normalized
variables (zero average and unit variance), the constants in the
problem (such as $f(1)$ and $f'_{\textsc{s}}(1)$) have no effect on
the distributions. Hence, in this case the {\textsc{pdf}}'s for the
two problems are identical. The sum over a finite volume, used to
compute the global mesoscopic $m$ does not modify this conclusion.  In
particular, in $2d$ the {\textsc{pdf}} can be approximated by a
generalized Gumbel distribution with parameter $a\approx \pi/2$ (see
App.~\ref{gumbel}), as discussed in detail in Ref.~\cite{Peter}.

We want to stress that the description of the {\textsc{pdf}}'s with
Gumbel-like extreme statistics is, in general, just an approximation
that correctly captures their main features, namely a single maximum,
negative skewness that crosses over to a positive skewness, and an
exponential or near exponential tail for small fluctuations below the
mean.  Moreover, $C_r$ is not {\it a priori} an extremal quantity so
there is no deep reason why it should be distributed following the
extreme value statistics.

\vspace{0.5cm}
\noindent
\underline{Case ${\cal C}\approx 0$}
\vspace{0.5cm}

If the global correlation is close to zero, $n=\Delta s/\delta s$ is
large. In this case, the sum in
Eq.~(\ref{discrete-nonlinear-collapse}) has many terms, and since all
terms in the sum are positive, the total will have some average that
is proportional to $\Delta s$ plus fluctuations. Some reflection point
us to the same situation that we presented in Sect.~\ref{sec:linear},
where we discussed the linear regime. Thus, we can write the local
correlations as
\begin{equation}
  C_r = f\left({\rm e}^{-\Delta s - \sqrt{\Delta s}\;X_r}\right) \;,
\end{equation}
where the variable $X_r$ is distributed according to
Eq.~(\ref{eq:X_r-dist}).  From here one can show that if $f(x)$ is a
monotonic increasing function, the {\textsc{pdf}} $P(C_r)$ is
positively skewed (as opposed to what is found when ${\cal C}\approx
q_{\textsc{ea}}$).  Indeed, expanding $C_r$ around $X_r\sim 0$,
\begin{eqnarray}
  C_r &\approx & f\left({\rm e}^{-\Delta s}\right) -\sqrt{\Delta
  s}\;X_r\;{\rm e}^{-\Delta s}\;f'\left({\rm e}^{-\Delta s}\right)+ \\ & &
  \frac{\Delta s}{2}\;X^2_r\; \left[{\rm e}^{-2\Delta
  s}\;f''\left({\rm e}^{-\Delta s}\right)+ {\rm e}^{-\Delta
  s}\;f'\left({\rm e}^{-\Delta s}\right) \right] \nonumber
\label{eq:exp-Clow}
\end{eqnarray}
one sees that the signs of the quadratic term in $X_r$ in
Eq.~(\ref{eq:exp-Clow}) and the quadratic term in $Y_r$ in
Eq.~(\ref{eq:expansion-C}) have opposite signs. Since this sign
dictates the skewness (recall that $X_r$ has a symmetric
distribution), one concludes that the {\textsc{pdf}} is positively
skewed in this limit, as also observed numerically and experimentally.

\subsection{Beyond a single scale}

The argument used above can be heuristically extended to deal with
more general problems. Let us assume that global correlations evolve
in a sequence of scales, each characterized by their own scaling
function, $h_k(t)$, and external function,
$f_k$,~\cite{Cugliandolo,Cuku2,Frme}
\begin{equation}
  C(t,t_w) = \sum_{k=1}^n f_k \left( \frac{h_k(t_w)}{h_k(t)}\right)
\end{equation}
with each $h_k(t)$ different monotonically growing
function of the original time.
This means
that when the ratio for a chosen value of $k$ varies as,
\begin{equation}
  0 \leq \frac{h_k(t_w)}{h_k(t)} \leq 1 \; ,
\end{equation}
all others are constant:
\begin{equation}
  \frac{h_{k' (< k)}(t_w)}{h_{k' (< k)}(t)} =0 \; ,
  \;\;\;\;\;\;\;\;\;\;\;\;\; \frac{h_{k' (> k)}(t_w)}{h_{k' (> k)}(t)}
  =1 \; \; .
\end{equation}
The boundary conditions are such that
\begin{eqnarray}
  \lim_{x\to 1^+}f_k(x) = q_k \;,
  \;\;\;\;\;\;\;\;\;\;
  \lim_{x\to 0}f_k(x) = 0 \; ,
\end{eqnarray}
with $0 \leq q_{k+1} < q_k \leq 1$, and $\sum_{k=1}^n q_k=1$.  In the
time regime in which the $k$-th ratio varies the global correlation
reads
\begin{equation}
C(t,t_w) = f_k\left(\frac{h_{k}(t_w)}{h_{k}(t)} \right) +
\sum_{k'> k} q_{k'}
\; .
\end{equation}
Motivated by what was presented in~\cite{Castilloetal,Chamonetal}
we propose that the local fluctuations can be parametrized as
\begin{eqnarray}
  C_r(t,t_w) &=& \sum_{k=1}^n f_k \left(
  \frac{{h_k}_r(t_w)}{{h_k}_r(t)}\right) \nonumber \\ &=& \sum_{k=1}^n
  f_k \left( {\rm e}^{{\phi_k}_r(t_w)-{\phi_k}_r(t)} \right) \; .
\end{eqnarray}
We assume that the local fluctuations are small in the sense that they
do not exchange scales. This is consistent with the supposition that
the Edwards-Anderson parameter, or more generally in this case the
plateau values $q_k$, are not modified by the fluctuations.

Based on the symmetry arguments already explained, we propose that
each ${\phi_k}_r(t)$ is distributed as in (\ref{S-effective}). This
leads, after the reparametrization of time, to
\begin{equation}
  s_k(t) \equiv \ln h_k(t)  \; ,
\end{equation}
and the change of variables $({\psi_k}_r(s_k))^2=\dot {\phi_k}_r(t)$,
to a Gaussian distribution for each ${\psi_k}_r(s_k)$,
\begin{equation}
  S^k_{\textsc{eff}} = K \int d^d r \int ds_k \;
  (\nabla{\psi_k}_r(s_k))^2 \; .
\end{equation}
Note that the $\dot {\phi_k}_r(s_k)$ are positive and we assumed that
the stiffness $K$ is $k$ independent, for simplicity.  This is simply
due to the fact that the $h_k(t)$ characterizing each two-times sector
are monotonically growing functions of time.

Thus, when we look at the connected correlations between the $\Delta
{\phi_k}_r|_{t_w}^t$, for given values of $t$ and $t_w$ that
correspond to the scale $k$ for the global correlation, only the
$k$-th scale makes a non-trivial (non-constant)contribution.  Since in
each two-time sector the global scaling of the {\textsc{pdf}} of local
correlations is the same as the one for the global ${\cal C}$, we find
that this result holds for all times. Once we have analyzed the case
of $n$ scales, we can take the limit $n\to\infty$ that corresponds to
the ultrametric scaling form and claim that in this case the
{\textsc{pdf}}'s scale as the global ones. Thus, we expect the
collapse of the type found for the facilitated models.

\subsection{Numerics for random surfaces}

In this section we numerically generate random surfaces with the
statistics given by (\ref{nonlinear-collapse}) and we calculate the
{\textsc{pdf}}'s of local correlations related to the surfaces using
Eq.~(\ref{local-corr-param}). In order to carry out numerical
evaluations of the local $C_r(t,t_w)$ {\textsc{pdf}}'s, let
\begin{equation}
\Delta\phi_r|^t_{t_w}=\delta s\;\sum_{i=1}^n\; \psi_r^2(s_i)
\; ,
\label{discrete-nonlinear-collapse2}
\end{equation}
where we have discretized the integral in Eq.~(\ref{nonlinear-collapse}) in
steps of the cutoff $\delta s$. We proceed as follows:

\begin{enumerate}

\item[{\it i.}] We generate independent Gaussian random surfaces
$\psi_r(s_i)$, which we choose to be of unit variance since the choice
of variance does not change the results qualitatively.  These surfaces
are generated with the distribution $P[\dot \psi_r(s_i)]\propto
{\rm e}^{- K \int d^dr  (\nabla \dot\psi_r)^2}$ (see
Eq.~(\ref{S-effective-psi})).

\item[{\it ii.}] We then sum the $\psi^2_r(s_i)$ over $n$
time slices. The number of slices determines the time difference
$t-t_w$.

\item[{\it iii.}] Next we generate the surfaces
$C_r(t,t_w)= {\rm e}^{-\Delta\phi_r|^t_{t_w}}$, {\it i.e.} we choose for
simplicity the ``external'' function $f(x)=x$ [see
Eq.~(\ref{global-C})].

\item[{\it iv.}]  Finally, we look at the {\textsc{pdf}}'s $P(C_r)$ of
the normalized coarse-grained local correlations $C_r$ (subtracting
the average, scaling by the standard deviation, and coarse-graining
over a box of volume $\ell^3=(2M+1)^3$) as a function of the number of
slices $n$.

\end{enumerate}

Before presenting the numerical results, let us discuss in more detail
the issue of coarse-graining the correlation $C_r$ over a certain
volume. There are actually two levels of coarse-graining for theories
defined on a lattice. First, in order to define continuous fields
$h_r(t), \phi_r(t)$, and ultimately $\psi_r(t)$, a coarse-graining of
a lattice theory has already occurred.  Coarse-graining the local
correlation $C_r$ over a given volume is a {\it second} level of
coarse-graining.  The connection to the {\textsc{xy}} model and the
fluctuations of the global magnetization studied by Bramwell {\it et
  al.}~\cite{Peter} is helpful in clarifying this point. Starting from
a lattice theory, one coarse-grains to obtain a Landau-type theory of
a continuous variable $m_r$. Then, to study the fluctuations of the
total magnetization of the system over a given volume, one must
average $m_r$ over a second coarse-graining volume (which in the case
of the global magnetization is the entire volume of the system). In
the numerical simulations we present here, we start from an action for
the $\psi_r$ that is Gaussian. The first level of coarse-graining that
led to an action for a continuous $\phi_r$ or $\psi_r$ was assumed to
have taken place already. So even though in the numerical calculations
we employ a discretization, one should keep in mind that a first
coarse-graining has been invoked.

\begin{figure}[ht]
\center{\includegraphics[scale=.68]{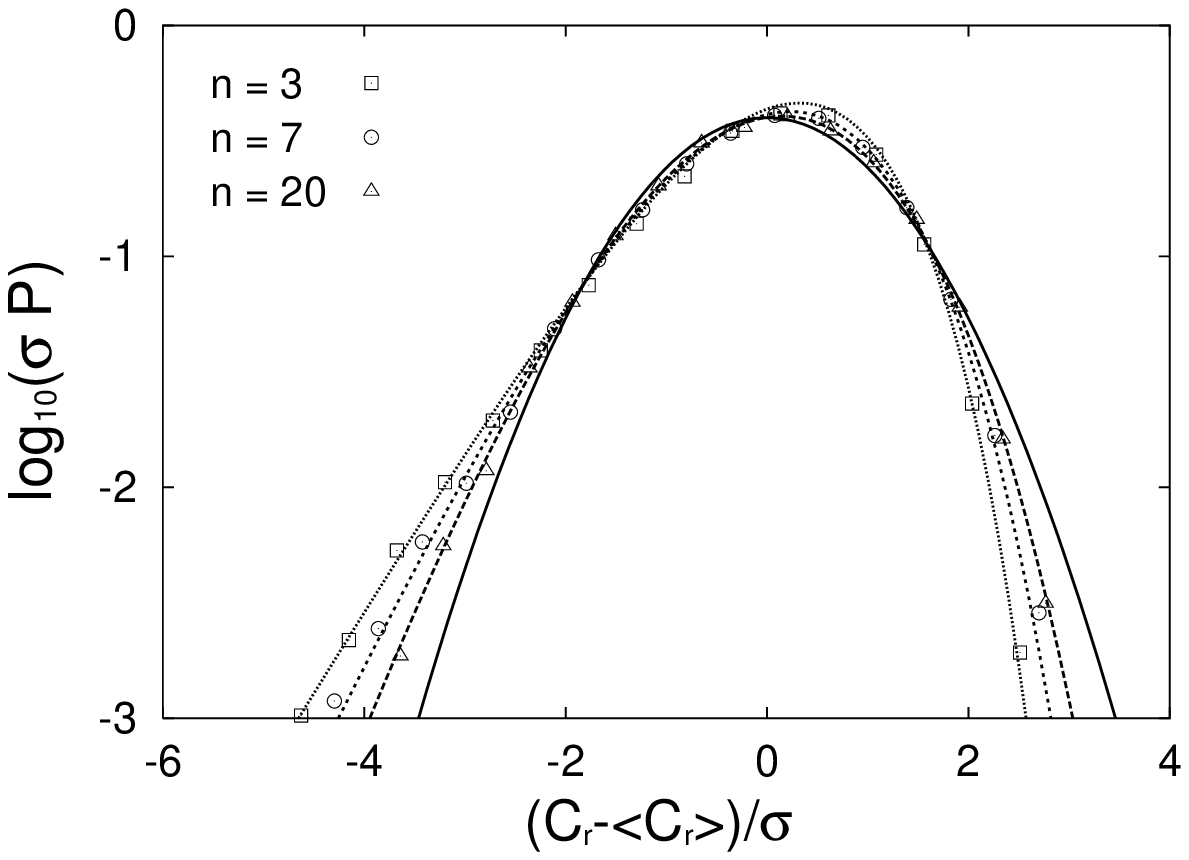}
\includegraphics[scale=.68]{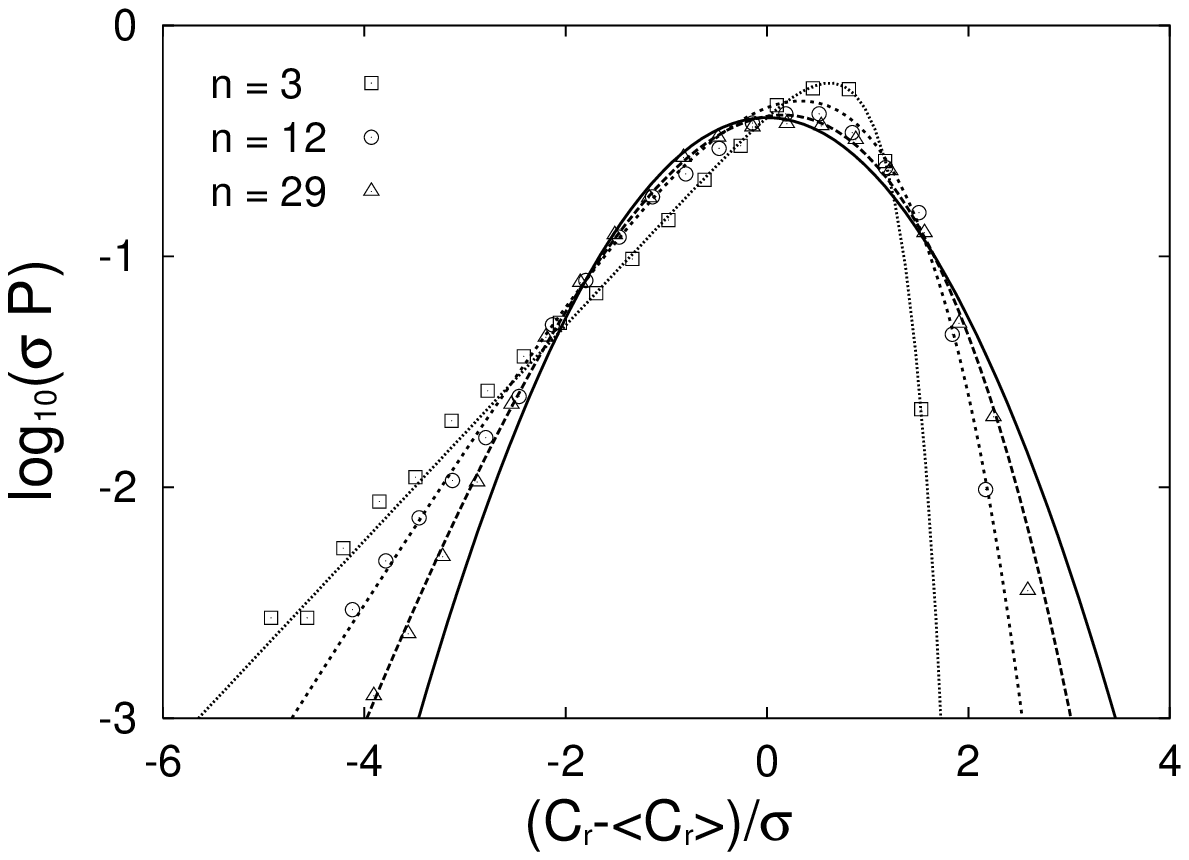}
}
\caption{Normalized {\textsc{pdf}}'s $P(C_r)$ for local coarse-grained
  correlations $C_r$ generated by the random surfaces suggested by the
  analytical theory. The curves shown are for sums over $n$ slices.
  As $n$ increases they become less skewed and approach the Gaussian
  distribution (full line). Top: $d=3$, system of linear size $L=32$,
  and the coarse-graining cell of linear size $\ell=3$. The Gumbel
  parameter is $a=2.2,\,4.5,\,11$ (for increasing $n$). Bottom: $d=2$, $L=32$,
  and $\ell=5$; $a=0.4,\,2,\,10$ (for increasing $n$).
  }
\label{fig:randsurface-GumbelM1}
\end{figure}

In Fig.~\ref{fig:randsurface-GumbelM1}, we show for a system of size
$L=32$ the dependence of the $P(C_r)$ on the number of slices.  We
show curves for $n=4,8,12,16,20$, and for a coarse-grained volume
$\ell=3$.  The smaller the value of $n$, the smaller is the ratio
$h_{\textsc{s}}(t)/h_{\textsc{s}}(t_w)$. Several comments are in
order. First, notice that for small $n$, the curves clearly deviate
from a Gaussian form and are reasonably well fit by a generalized
Gumbel distribution (see App.~A).  The curves become more Gaussian as
$n$ increases. This behavior is qualitatively similar to the one we
obtained for the facilitated spin model and the $3d$ {\textsc{ea}}
model we studied numerically in Sect.~\ref{fluctuations} as well as
the experimental observations of Cipelletti {\it et al.}~\cite{Luca2}.
The approach to a Gaussian behavior is however very slow and we do not
reach the Gaussian form with the values of $n$ used.

\subsection{Essential ingredients for universal fluctuations}

Why do the analytical arguments we presented above, based on a
symmetry (reparametrization invariance), lead to the universal
behavior and collapse (for a fixed global correlation) of the $P(C_r)$
that is observed in the facilitated spin models, in the $3d$
Edwards-Anderson model, and in some experiments~\cite{Luca2}?  We can
single out two essential elements of the theoretical description that
allow it to account for this behavior.

The first ingredient is that the theory proposed in
Refs.~\cite{Castilloetal,Castilloetal2,Chamonetal} contains a massless
Goldstone mode associated with the {\it spatial} fluctuations in the
asymptotic long-time (and long-time-difference) limit.  This means
that correlation functions such as those in Eq.~(\ref{N-corr}) are
described by algebraically decaying functions ${\cal F}({\bf r}_1,{\bf
  r}_2,\dots,{\bf r}_N)$. The absence of a correlation length $\xi$
leads to the non-Gaussian behavior for quantities averaged over a
finite volume, similarly to the studies in Refs.~\cite{Racz,Peter}. Of
course, one expects that universal behavior of the {\textsc{pdf}}'s
will occur only after coarse-graining. Note however that it is the
absence of a correlation scale that implies that the law of large
numbers will not take effect as the box size increases. So the theory
based on reparametrization invariance of the action gives rise to a
non-Gaussian distribution for coarse-grained $C_r(t,t_w)$ (in addition
to any coarse-graining needed to define a theory in the space
continuum) because of the algebraic correlations originating from the
massless Goldstone mode. In the theory of
Refs.~\cite{Castilloetal,Castilloetal2,Chamonetal} we argued that the
correlation scale $\xi\to \infty$ in the asymptotic long-time limit.
However, for finite times there is a finite
$\xi$~\cite{Castilloetal2}. Coarse-graining much beyond this length
scale leads to Gaussian distributions.

The second element that the analytical theory contains is that all
dependencies on the times are functions of $\Delta s=s(t)-s(t_w)$.
This leads to the collapse of distributions (not only their bulk
averages) as a function of $\ln
h_{\textsc{s}}(t)/h_{\textsc{s}}(t_w)$. In addition, the fact that the
different $n=\Delta s/\delta s$ slice contributions
$\dot\psi_r^2(s_i)$ are uncorrelated, and that they enter in the local
correlations $C_r(t,t_w)$ through the sum over all slices, is the
reason why the skewness of the distributions depend on $n$, as shown
in Fig.~\ref{fig:randsurface-GumbelM1}.  Again, this is in good accord
with the experimental observations of Ref.~\cite{Luca2} and with the
numerical results for the models that we study in this paper.

In summary, we believe that the qualitative features shared by the
{\textsc{pdf}}'s of local correlators in the facilitated models, in
the $3d$ {\textsc{ea}} model, and in the experimental observations of
Cipelletti {\it et al.}~\cite{Luca2} can be understood as a
consequence of certain spatial and temporal correlations of the
$C_r(t,t_w)$, which are contained in the theory based on
reparametrization invariance.

\section{Conclusions}
\label{conclusions}

In equilibrium and away from criticality any global observable of a
macroscopic system has Gaussian fluctuations. At criticality, instead,
one observes non-Gaussian fluctuations due to the divergence of the
correlation length and the non applicability of the central limit
theorem. Still, scale invariance at the critical point constrains the
possible probability distributions; these are determined by the
universality class to which the systems belong.  A similar criterion
to classify the probability distributions of the fluctuations of
macroscopic observables in critical nonequilibrium systems is based on
the use of symmetries. It has been proposed in~\cite{Racz}.

The glass transition, where the system falls out of equilibrium, is a
dynamic crossover and neither a dynamic nor a thermodynamic
transition. However, many features of glassy dynamics resemble
critical relaxation such as the fact that correlations do not decay
exponentially, but with much slower decay forms. One could expect then
that some concepts that have been useful to study critical phenomena
could also apply to the glassy dynamics~\cite{Whitelam,Biroli}.

In this paper we continued the study of local fluctuations in the
dynamics of glassy systems. On the theoretical side we improved the
sigma model, or random manifold action, proposed
in~\cite{Castilloetal,Castilloetal2,Chamonetal} to capture not only
the scaling in time of the {\textsc{pdf}}'s but also their functional
form.  On the numerical side, we showed that part of the predictions
of the theory tested in~\cite{Castilloetal,Castilloetal2} using the
$3d$ Edwards-Anderson model also hold for a non-disordered kinetically
constrained spin model of a glass.  This confirms that the existence
of quenched disorder is not important in this respect and that kinetic
frustration and energetic frustration lead to very similar
non-equilibrium dynamics.  Moreover, we tested the evolution of the
form of the {\textsc{pdf}}'s for these two models and verified that
they behave in a way that resembles strongly the experimental
observations in~\cite{Luca2}.

One could expect that a more detailed and extended analysis of the
local dynamics of glassy systems might lead to better differentiation
between models and systems. For instance, one could find that the
simple action (\ref{nonlinear-collapse}) is not enough to reproduce
the special form of {\textsc{pdf}}'s found in a system and hence be
forced to include other terms that we here neglected.

\bigskip

This work was supported in part by the NSF grants DMR-0305482 and
INT-0128922 (CC), a CNRS-NSF collaboration grant, an ACI Jeunes
Chercheurs ``Algorithmes d'opti\-mi\-sation et sys\-t\`emes
d\'e\-sor\-don\-n\'es quan\-tiques'', the Guggenheim Foundation
(LFC), FQRNT (PC),
and the NSF grants PHY99-07949 (CC, LFC, DR) and CHE-0134969 (PC, DR).

CC, LFC, and DR thank the KITP Santa Barbara, and LFC the Condensed
Matter group at the Abdus Salam International Centre for Theoretical
Physics for their hospitality
during part of the preparation of this work.  We especially thank
H.~E.~Castillo, S.~Franz, M.~P.~Kennett and J.~Polchinski for very
useful discussions.

\appendix

\section{Generalized Gumbel distributions}
\label{gumbel}

The Gumbel distribution is defined as
\begin{eqnarray}
\label{eq:GumbelA}
\Phi_a(y) & = & \frac{|\alpha| a^a}{\Gamma(a)}
\; {\rm e}^{a\left(\alpha(y-y_0)-{\rm e}^{\alpha(y-y_0)}\right)} \,,
\label{Gumbel-dist}
\end{eqnarray}
where $\Gamma(a)$ is the {\it gamma} function.  The parameters $y_0$
and $\alpha$ control the position of the center and the width of the
distribution, respectively. One could define standard distributions by
requiring that the center is at zero, and the width is unity. These
conditions fix, for a given $a$, $y_0(a)$ and $\alpha(a)$ as
\begin{eqnarray}
\alpha = \sqrt{\Psi'(a)} \;, \;\;\;\;\;\;\;\;\;\;
\alpha y_0  = \log a - \Psi(a) \,,
\label{Gumbel-coefficient}
\end{eqnarray}
where $\Psi(a)$ is the {\it digamma} function
$\Psi(a)=\Gamma'(a)/\Gamma(a)$.

The Gumbel distribution with $a=1$ appears as one kind of asymptotic
extreme value statistics when searching for the distribution of the
maximum (or the minimum) of a sequence of independent identically
distributed random variables with a probability density decaying
faster than any power law. Extensions with integer parameter $a>1$ are
the results of searching for the distribution of the $a$-th largest
(smallest) value in the sequence.  For any $a$, there is a choice of
sign for $\alpha$ and the two signs correspond to the Gumbel
statistics of either extreme minima or extreme maxima. Further
extensions with non-integer parameter $a<1$ have been found when the
elements of the sequence are correlated over a distance
$1/a$~\cite{Gumbel}.  For $a=\pi/2$ one recovers the distribution
studied by Bramwell, Holdsworth and Pinton~\cite{BHP}.

Finally, let us study two limits of the generalized Gumbel-like
distribution (\ref{Gumbel-dist}).  A Gaussian distribution of unit
variance is obtained by taking the $a\to\infty$ limit of $\Phi_a(y)$
while holding the product $\alpha^2 a=1$ fixed:
\begin{widetext}
\begin{eqnarray}
\lim_{a\to\infty,\alpha^2a=1}\;
\Phi_a(y)&=& \lim_{a\to\infty,\alpha^2a=1}\;
\frac{a^a}{\sqrt{a}\;\Gamma(a)}
\; {\rm e}^{a\left(\alpha(y-y_0)-1-\alpha(y-y_0)-\frac{1}{2}\alpha^2(y-y_0)^2+
{\cal O}(\alpha^3)\right)}\nonumber\\
&=&\lim_{a\to\infty} \;
\frac{a^a {\rm e}^{-a}}{\sqrt{a}\;\Gamma(a)}
\; {\rm e}^{-\frac{1}{2}(y-y_0)^2}\nonumber\\
&=&
\frac{1}{\sqrt{2\pi}}
\;{\rm e}^{-\frac{1}{2}(y-y_0)^2}
\ ,
\end{eqnarray}
\end{widetext}
where in the last step we used Stirling's formula to get the
normalization factor.

Another interesting limit is when $a\to 0$ while holding $|\alpha| a=1$:
\begin{widetext}
\begin{eqnarray}
\lim_{a\to0,|\alpha| a=1}\;
\Phi_a(y) &=& \lim_{a\to0}\;
\frac{a^a}{a\Gamma(a)}
\;{\rm e}^{{\rm sgn}\alpha\;(y-y_0)} \;\theta\left(-{\rm sgn}\alpha\;(y-y_0)\right)
 \nonumber \\
&=&
\;{\rm e}^{{\rm sgn} \alpha\;(y-y_0)}
\;\theta\left(-{\rm sgn}\alpha\;(y-y_0)\right)\,
\end{eqnarray}
\end{widetext}
where we used that $\lim_{a\to 0} a\Gamma(a)=\Gamma(1)=1$. Thus, the
$a\to 0$ limit leads to exponential distributions either starting or
ending at $y=y_0$.

\section{Intrinsic skewness of uncorrelated dynamics}
\label{uncorrelated}
\label{app:toy}

The probability distribution of local correlation functions evaluated
at finite times and coarse-grained over a sufficiently large cells is,
to a first approximation, normally distributed around the bulk value.
However, for the times and cell sizes considered in this paper, one
needs to consider how the intrinsic distribution of unconstrained
spins contributes to the skewing of the distribution of local
correlations.  In this Appendix, we consider the degree of intrinsic
skewing that arises from uncorrelated dynamics on a lattice.

To differentiate between correlated and intrinsic skewness, we look at
a model of paramagnetism without any facilitation.  The Hamiltonian is
then the same as in the \textsc{fa} model.
This toy model has no dependence on
the dimensionality or the neighboring environment, and depends only on
the number of spins included in the cells. Therefore, all sites are
independent and the probability $p$ that an initially up spin is down
at time $t$ is the same everywhere ($q$ is the equivalent probability
for the spins initially down). Using Metropolis Monte Carlo single
spin flips, one finds the following distribution for a cell of $N$
spins using binomial distribution arguments,
\begin{widetext}
\begin{eqnarray} \label{eq:statweigh}
  \mathcal{P}(n,d,u) &=& \frac{N!}{(n-d)!d!(N-n-u)!u!} \,\, c^n
  (1-c)^{N-n} \, p^d (1-p)^{n-d} \, q^u (1-q)^{N-n-u} \,.
\end{eqnarray}
\end{widetext}
Here $n$ is the number of spins up initially up, $c$ is the initial
bulk concentration of up spins, $d$ the number of initially up spins
that are down at time $t$, and $u$ of initially down spins which are
up at time $t$. In the equilibrium case, $p$ and $q$ are related by
\begin{eqnarray}
q_{eq}=c_{eq}/(1-c_{eq})p_{eq}
\end{eqnarray}
The same relationship (without the subscripts) holds for a constant
concentration system, but not in the case of a steadily decreasing
concentration, as in the quenched facilitated models.  That however
can be resolved by including a scaling factor $X$ proportional to the
concentration during the quench as a function of $t_w$ and $t$.

For a given ensemble average value ${\cal C}$ of the single site
correlation, one can obtain the {\textsc{pdf}} for the correlation
averaged over a number $V_r=\ell^d$ of uncorrelated spins. (Notice
that, since these are free spins without any dynamical constraint, the
number of time steps necessary to reach the average value ${\cal C}$
is much smaller than that for the kinetically constrained models.)
Using the result in (\ref{eq:statweigh}), one can obtain the
corresponding value of the cell connected correlation function
\begin{eqnarray} \label{eq:toycorrel}
  C_{c,norm}(t)\equiv \overline{c(0)c(t)}
  -\overline{c(0)}\times\overline{c(t)}
\end{eqnarray}
where $\overline{c(t)}=\frac{n-d-u}{N}$,
$\overline{c(0)}=\frac{n}{N}$, and
$\overline{c(0)c(t)}=\frac{n-d}{N}$. The overline refers to the
sub-ensemble average.

Using bulk values for the different parameters, the statistical
component of the distribution of the local spin correlation can be
obtained, by numerically evaluating Eq.~(\ref{eq:toycorrel}) for all
$n$, $u$, and $d$ and properly weighting them according to
Eq.~(\ref{eq:statweigh}).  Tables~\ref{tab:param1}, \ref{tab:param2},
and \ref{tab:param3} show the parameters used in the different models.
In the case of the plaquette models, the same discussion applies, but
with a null magnetic field instead.

\vspace{0.25cm}
\begin{table}

\begin{center}
\begin{tabular}{|c|c|c|c|c|c|}
\hline
$t_w$& $10^1$ & $10^2$ & $10^3$ & $10^4$ &$10^5$ \\ $t$ & $30$ & $45$
& $90$ & $330$ &$1000$ \\\hline $c$ & 0.30865 & 0.27257 & 0.21203 &
0.15505 & 0.1685 \\ $X$ & 0.938 & 0.969 & 0.988 & 0.995 & 0.9995 \\
$p$ & 0.233 & 0.230 & 0.237 & 0.258 & 0.264 \\ $a$ & 100 & $\sim\infty$
& $\sim\infty$ & $\sim\infty$ & $\sim\infty$ \\\hline
\end{tabular}
\end{center}
\caption{Unconstrained parameters for a decay of the global
correlations to a value of $0.7$ of the initial value for the
\textsc{fa}-$2d2n$ model for a quenching $T=0.6$ and a box size of
$N=10\times10$. The parameters are the $t_w$ defect concentration $c$,
the concentration scaling factor $X$ for time $t$, and the probability
$p$ of defect disappearance at time $t$. The resulting approximative
Gumbel $a$ parameter is also given.} \label{tab:param1}
\end{table}

\begin{table}

\begin{center}
\begin{tabular}{|c|c|c|c|c|c|c|c|}
\hline
$t_w$& $10^1$ & $10^2$ & $10^3$ & $10^4$ &$10^5$ \\ $t$ & $15$ & $26$
& $81$ & $370$ &$2300$ \\\hline $c$ & 0.3299 & 0.2903 & 0.2399 &
0.2007 & 0.1685 \\ $X$ & 0.959 & 0.982 & 0.993 & 0.996 & 0.998 \\ $p$
& 0.220 & 0.219 & 0.2305 & 0.241 & 0.249 \\ $a$ & 80 & 80 & 100 &
$\sim\infty$ & $\sim\infty$ \\\hline
\end{tabular}
\end{center}
\caption{Unconstrained parameters for a decay of the global
correlations to a value of $0.7$ of the initial value for the
\textsc{fa}-$3d3n$ model for a quenching $T=0.8$ and a box size of
$N=5\times5\times5$.  Parameter definitions are identical to those in
Table~\ref{tab:param1}.} \label{tab:param2}
\end{table}

\begin{table}

\begin{center}
\begin{tabular}{|c|c|c|c|c|c|c|c|}
\hline
$t_w$& $10^1$ & $10^2$ & $10^3$ & $10^4$ &$10^5$ \\ $t$ & $2$ & $4$ &
$10$ & $32$ &$100$ \\\hline $c$ & 0.3299 & 0.2903 & 0.2399 & 0.2007 &
0.1685 \\ $X$ & 0.992 & 0.997 & 0.9996 & 1.000 & 1.000 \\ $p$ & 0.0812
& 0.0908 & 0.0902 & 0.0962 & 0.0993 \\ $a$ & 15 & 20 & 40 & 90 & 100
\\ \hline
\end{tabular}
\end{center}
\caption{Unconstrained parameters for a decay of the global
correlations to a value of $0.88$ of the initial value for the
\textsc{fa}-$3d3n$ model for a quenching $T=0.8$ and a box size of
$N=5\times5\times5$.  Parameter definitions are identical to those in
Table~\ref{tab:param1}.} \label{tab:param3}
\end{table}

In the first two cases (Table~\ref{tab:param1} and \ref{tab:param2})
the bulk correlation is chosen to have a rather large value and the
Gumbel parameters found are extremely large corresponding to
approximately Gaussian distributions. No interesting effect is
observed in these cases. In the last example considered
(Table~\ref{tab:param3}) the bulk correlation is chosen to take a
smaller value and the parameters found indicate that the distribution
is skewed. However, if one compares to the results obtained for the
same model with facilitation one concludes that the non-Gaussian
effect is here much weaker and tends to disappear rather quickly when
the times involved get long.

In summary, the simple correlations that follow from coarse-graining
of some binomially distributed single-site spin-spin correlation whose
average is restricted to be ${\cal C}$ cannot account for the
{\textsc{pdf}}s we find for the kinetically constrained model. Spatial
correlations between the spin variables are needed to account for both
the width and skeweness of the distributions.


\begin{thebibliography}{99}

\bibitem{review-glasses} M.~D.~Ediger, C.~A.~Angell, and S.~R.~Nagel,
{\em J.~Phys. Chem.} {\bf 100}, 13200 (1996).

\bibitem{Ediger} H.~Sillescu, {\em J.~Non-Crystal. Solids} {\bf 243},
81 (1999); M.~D.~Ediger, {\em Annu. Rev. Phys. Chem.} {\bf 51}, 99
(2000).

\bibitem{Kegel} W.~K.~Kegel and A.~V.~Blaaderen, {\em Science} {\bf
287}, 290 (2000).

\bibitem{Weitz} E.~Weeks, J.~C.~Crocker, A.~C.~Levitt, A.~Schofield,
and D.~A.~Weitz, {\em Science} {\bf 287}, 627 (2000).

\bibitem{Weeks1} E.~R.~Weeks and D.~A.~Weitz, {\em Phys. Rev. Lett.}
{\bf 89}, 095704 (2002).

\bibitem{Weeks2} R.~E.~Courtland and E.~R.~Weeks, {\em J.~Phys.:
Condens. Matter} {\bf 15}, S359 (2003).

\bibitem{Nathan} E.~Vidal~Russell, N.~E.~Israeloff, L.~E.~Walther, and
H.~Alvarez Gomariz, {\em Phys. Rev. Lett.}{ \bf 81}, 1461 (1998);
L.~E.~Walther, N.~E.~Israeloff, E.~Vidal Russell, and H.~Alvarez
Gomariz, {\em Phys. Rev.~B} {\bf 57}, R15112 (1998); E.~Vidal Russel
and N.~E.~Israeloff, {\em Nature (London)} {\bf 408}, 695 (2000).

\bibitem{Miller} R.~S.~Miller and R.~A.~MacPhail, {\em J.~Phys. Chem.}
  {\bf 101}, 8635 (1997).

\bibitem{Sergio2} L.~Buisson, L.~Bellon, and S.~Ciliberto,
 cond-mat/0210490, to appear in Proceedings of ``III Workshop on
Non-Equilibrium Phenomena" (Pisa 2002).

\bibitem{Luca2} L.~Cipelletti, H.~Bissig, V.~Trappe, P.~Ballestat, and
S.~Mazoyer, {\em J.~Phys.: Condens. Matter} {\bf 15}, S257 (2003);
H.~Bissig, V.~Trappe, S.~Romer, and L.~Cipelletti,  cond-mat/0301265.

\bibitem{Laird} B.~B.~Laird and H.~R.~Schober, {\em Phys. Rev. Lett.}
{\bf 66}, 636 (1991); H.~R.~Schober and B.~B.~Laird, {\em
Phys. Rev.~B} {\bf 44}, 6746 (1991).

\bibitem{2d} T.~Muranaka and Y.~Hiwatari, {\em Phys. Rev.~E} {\bf 51},
R2735 (1995); M.~M.~Hurley and P.~Harrowell, {\em Phys. Rev.~E} {\bf
52}, 1694 (1995); D.~N.~Perera and P.~Harrowell, {\em Phys. Rev.~E}
{\bf 59}, 5721 (1999).

\bibitem{Donati99} W.~Kob, C.~Donati, S.~J.~Plimpton, P.~H.~Poole, and
S.~C.~Glotzer, {\em Phys. Rev. Lett.} {\bf 79}, 2827 (1997);
C.~Donati, J.~F.~Douglas, W.~Kob, S.~J.~Plimpton, P.~H.~Poole, and
S.~C.~Glotzer, {\em Phys. Rev. Lett.} {\bf 80}, 2338 (1998);
C.~Donati, S.~C.~Glotzer, P.~H.~Poole, W.~Kob, and S.~J.~Plimpton,
{\em Phys. Rev.~E} {\bf 60}, 3107 (1999).

\bibitem{Heuer} A.~Heuer and A.~Okun, {\em J.~Chem. Phys.} {\bf 106},
6176 (1997); B.~Doliwa and A.~Heuer, {\em Phys. Rev. Lett.} {\bf 80},
4915 (1998).

\bibitem{Johnson} G.~Johnson, A.~I.~Mel'cuk, H.~Gould, W.~Klein, and
R.~D.~Mountain, {\em Phys. Rev.~E} {\bf 57}, 5707 (1998).

\bibitem{Yamamoto} R.~Yamamoto and A.~Onuki, {\em Phys. Rev.~E} {\bf
58}, 3515 (1998).

\bibitem{Starr02} F.~W.~Starr, S.~Sastry, J.~F.~Douglas, and
S.~C.~Glotzer, {\em Phys. Rev. Lett.} {\bf 89}, 125501 (2002).

\bibitem{Ogli} C.~Oglichler and H.~R.~Schober, {\em Phys. Rev.~B} {\bf
59}, 811 (1999).

\bibitem{Vollmayr} K.~Vollmayr-Lee, W.~Kob, K.~Binder, and
A.~Zippelius, {\em J.~Chem. Phys.} {\bf 116}, 5158 (2002).

\bibitem{Castilloetal} H.~Castillo, C.~Chamon, L.~F.~Cugliandolo, and
M.~P.~Kennett, {\em Phys. Rev. Lett.} {\bf 88}, 237201 (2002).

\bibitem{Castilloetal2} H.~Castillo, C.~Chamon, L.~F.~Cugliandolo,
J.~L.~Iguain, and M.~P.~Kennett, {\em Phys. Rev.~B} {\bf 68}, 134442
(2003).

\bibitem{review-Bouchaud} J.-P.~Bouchaud, L.~F.~Cugliandolo,
J.~Kurchan, and M.~M\'ezard, in {\it Spin-Glasses and Random Fields},
A.~P.~Young ed. (World Scientific, 1998).

\bibitem{Cugliandolo} L.~F.~Cugliandolo, in {\em Slow Relaxations and
Nonequilibrium Dynamics in Condensed Matter}, J.-L.~Barrat {\em et
al.} eds. (Springer-Verlag, 2002),  cond-mat/0210312.

\bibitem{Cuku93} L.~F.~Cugliandolo and J.~Kurchan, {\em
Phys. Rev. Lett.}  {\bf 71}, 173 (1993).

\bibitem{Bocukume} J.-P.~Bouchaud, L.~F.~Cugliandolo, J.~Kurchan, and
M.~M\'ezard, {\em Physica A} {\bf 226}, 243 (1996).

\bibitem{free-energy} J.~Kurchan and L.~Laloux, {\em J.~Phys.~A} {\bf
29}, 1929 (1996); A.~Cavagna, {\em Europhys. Lett.} {\bf 53}, 490
(2001).

\bibitem{inherent} L.~Angelani, G.~Ruocco, M.~Sampoli, and
F.~Sciortino, {\em J.~Chem. Phys.} {\bf 119}, 2120 (2003); A.~Crisanti
and F.~Ritort, {\em Europhys. Lett.} {\bf 52}, 640 (2000).

\bibitem{FA} G.~H.~Frederickson and H.~C.~Andersen, {\em
Phys. Rev. Lett.} {\bf 53}, 1244 (1984); G.~H.~Frederickson and
H.~C.~Andersen, {\em J.~Chem. Phys.} {\bf 83}, 5822 (1985);
G.~H.~Fredrickson and S.~A.~Brawer, {\em J.~Chem. Phys.} {\bf 84},
3351 (1986).

\bibitem{KoAn} W.~Kob and H.~C.~Andersen, {\em Phys. Rev.~E} {\bf 48},
4364 (1993).

\bibitem{JaKr} J.~J\"{a}ckle and A.~Kr\"{o}nig, {\em J.~Phys.:
Condens. Matter} {\bf 6}, 7633 (1994); J.~J\"{a}ckle, {\em J.~Phys.:
Condens. Matter} {\bf 14}, 1423 (2002).

\bibitem{aging-kinetic} J.~Kurchan, L.~Peliti, and M.~Sellitto, {\em
Europhys. Lett.} {\bf 39}, 365 (1997); M.~Sellitto, {\em
Euro. Phys.~J.~B} {\bf 4}, 135 (1998); M.~Sellitto, {\em J.~Phys.:
Condens. Matter} {\bf 14}, 1455 (2002).

\bibitem{reviews-kinetic} F.~Ritort and P.~Sollich, {\em Adv. Phys.}
{\bf 52}, 219 (2003).

\bibitem{Harrowell} S.~Butler and P.~Harrowell, {\em J.~Chem. Phys.}
{\bf 95}, 4454 (1991); S.~Butler and P.~Harrowell, {\em
J.~Chem. Phys.} {\bf 95}, 4466 (1991); P.~Harrowell, {\em
Phys. Rev.~E} {\bf 48}, 4359 (1993); M.~Foley and P.~Harrowell, {\em
J.~Chem. Phys.} {\bf 98}, 5069 (1993).

\bibitem{Chandler} J.~P.~Garrahan and D.~Chandler, {\em
Phys. Rev. Lett.} {\bf 89}, 035704 (2002); L.~Berthier and
J.~P.~Garrahan, {\em Phys. Rev.~E} {\bf 68}, 041201 (2003);
J.~P.~Garrahan and D.~Chandler, {\em Proc. Natl. Acad. Sci. USA} {\bf
100}, 9710 (2003).

\bibitem{Chamonetal} C.~Chamon, M.~P.~Kennett, H.~E.~Castillo, and
L.~F.~Cugliandolo, {\em Phys. Rev. Lett.} {\bf 89}, 217201 (2002).

\bibitem{book} N.~D.~Goldenfeld, {\em Lectures on Phase Transitions
and the Renormalisation Group} (Addison-Wesley, Reading Mass.,1992).

\bibitem{Racz} Z.~R\'acz, ``Nonequilibrium phase transitions'', in
{\em Slow Relaxations and Nonequilibrium Dynamics in Condensed
Matter}, J.-L.~Barrat {\it et al.} eds. (Springer-Verlag, 2002);
G. Gy\"{o}gyi, P.C.W. Holdsworth, B. Portelli and Z. R\'acz, {\em Phys
Rev. E} {\bf 68}, 056116 (2003); T.~Antal, M.~Droz, G.~Gy\"orgy, and
Z.~R\'acz, {\em Phys. Rev. Lett.} {\bf 89}, 240601 (2001); G.~Foltin,
K.~Oerding, Z.~R\'acz, R.~L.~Workman, and R.~K.~P.~Zia, {\it
Phys. Rev.~E} {\bf 50}, 639 (1994).

\bibitem{Peter} S.~T.~Bramwell, J.~Y.~Fortin, P.~C.~W.~Holdsworth,
S.~Peysson, J.~F.~Pinton, B.~Portelli, and M.~Sellitto,
{\em Phys. Rev.~E} {\bf 63}, 041106 (2001); B.~Portelli,
P.~C.~W.~Holdsworth, M.~Sellitto, and S.~T.~Bramwell,
{\em Phys. Rev.~E} {\bf 64}, 036111 (2001).

\bibitem{Sharon} N.~La\v{c}evi\'c, F.~W.~Starr, T.~B.~Schroeder, and
S.~C.~Glotzer, {\em J.~Chem. Phys.} {\bf 119}, 7372 (2003);
C. Bennemann, C. Donati, J. Baschnagel and S. Glotzer, {\em
Nature} {\bf 399}, 246 (1999).

\bibitem{Silvio} S.~Franz, C.~Donati, G.~Parisi and S.~C.~Glotzer,
{\em Philos. Mag.~B} {\bf 79}, 1827 (1999). S. Franz and G. Parisi
{\em J.~Phys.: Condens. Matter} {\bf 12}, 6335 (2000).

\bibitem{Whitelam} S.~Whitelam, L.~Berthier, and J.~P.~Garrahan,
 cond-mat/0310207.
 
\bibitem{Biroli} G.~Biroli and J.-P.~Bouchaud,
 cond-mat/0401260.

\bibitem{Graham} I.~S.~Graham, L.~Pich\'e, and M.~Grant, {\em
Phys. Rev.~E} {\bf 55}, 2132 (1997).

\bibitem{EA} S.~F.~Edwards and P.~W.~Anderson, {\em J.~Phys.~F} {\bf
5}, 965 (1975).

\bibitem{Didier} D.~H\'erisson and M.~Ocio, {\em Phys. Rev. Lett.}
{\bf 88}, 257202 (2002).

\bibitem{Bonn} B.~Abou, D.~Bonn, and J.~Meunier, {\em Phys. Rev.~E }
{\bf 64}, 021510 (2001); D.~Bonn, S.~Tanase, B.~Abou, H.~Tanaka, and
J.~Meunier, {\em Phys. Rev. Lett.} {\bf 89}, 015701 (2002).

\bibitem{Virgile} A.~Knaebel, M.~Bellour, J.-P.~Munch, V.~Viasnoff,
F.~Lequeux, and J.~L.~Harden, {\em Europhys. Lett.} {\bf 52}, 73
(2000); V.~Viasnoff and F.~Lequeux, {\em Phys. Rev. Lett.} {\bf 89},
065701 (2002).

\bibitem{Luca1} L.~Cipelletti, S.~Manley, R.~C.~Ball, and D.~A.~Weitz,
{\em Phys. Rev. Lett.} {\bf 84}, 2275 (2000); L.~Ramos and
L.~Cipelletti, {\em Phys. Rev. Lett.} {\bf 87}, 245503 (2001).

\bibitem{Pouliquen} O.~Pouliquen, M.~Belzons, and M.~Nicolas,
 cond-mat/0305659.

\bibitem{Cuku2} L.~F.~Cugliandolo and J.~Kurchan, {\em J.~Phys.~A}
{\bf 27}, 5749 (1994).

\bibitem{Marco} M.~Picco, F.~Ritort, and F.~Ricci-Tersenghi,
{\em Eur. Phys.~J.~B} {\bf 21}, 211 (2001).

\bibitem{Mauro} G.~Biroli, S.~Franz, M.~Sellitto, and C.~Toninelli, in
preparation.

\bibitem{Cristina}
C.~Toninelli, G.~Biroli, and D.~S.~Fisher,  cond-mat/0306746.

\bibitem{review-spinglass} G.~Parisi, in {\em Slow Relaxations and
Nonequilibrium Dynamics in Condensed Matter}, J.-L.~Barrat {\it et
al.}  eds. (Springer-Verlag, 2002); D.~S.~Fisher, in {\em Slow
Relaxations and Nonequilibrium Dynamics in Condensed Matter},
J.-L.~Barrat {\it et al.} eds. (Springer-Verlag, 2002).

\bibitem{Bertin} E.~Bertin and J.-P.~Bouchaud, {\em J.~Phys.~A} {\bf
  35}, 3039 (2002).

\bibitem{East-model} J. J\"ackle and S.~Eisinger, {\em Z. Phys. B} {\bf
  84} 115 (1991); P.~Sollich and M.R.~Evans, {\em Phys. Rev. Lett.}
  {\bf 83} 3238 (1999); A.~Crisanti, F.~Ritort, A.~Rocco and
  M.~Sellitto, {\em J.~Chem. Phys.} {\bf 113} 10615 (2000); D. Aldous
  and P. Diaconis, {\em J. Stat. Phys.} {\bf 107} 945 (2002).

\bibitem{Frme} S.~Franz and M.~M\'ezard, {\em Europhys. Lett.} {\bf
 26}, 209 (1994); S.~Franz and M.~M\'ezard, {\em Physica A} {\bf 210},
 48 (1994).

\bibitem{Gumbel} E.~J.~Gumbel, {\it Statistics of Extremes} (Columbia
  University Press, New York, 1958).

\bibitem{BHP} S.~Bramwell, P.~W.~Holdsworth and J.-F.~Pinton, {\em
Nature} {\bf 396}, 552 (1998); S.~T.~Bramwell, J.~Y.~Fortin,
P.~C.~W.~Holdsworth, H.~J.~Jensen, S.~Lise, J.~M.~L\'opez,
M.~Nicodemi, J.~F.~Pinton, M.~Sellitto, {\em Phys. Rev. Lett.} {\bf
84}, 3744 (2000).

\bibitem{footnote} Note that $\dot \phi_r(s)$ is a positive definite
quantity; indeed, $0\leq \dot \phi_r(t) = d\phi_r(t(s))/ds \; ds/dt =
\dot \phi_r(s) \; \dot h_{\textsc{s}}(t)/h_{\textsc{s}}(t)$ implies
$\dot \phi_r(s) \geq 0$.

\end{thebibliography}
\end{document}